\documentclass[aps,prb,twocolumn,oneside,floatfix,sort&compress, superscriptaddress,showpacs,nobalancelastpage]{revtex4-1}

\usepackage{graphicx,xcolor}
\usepackage[utf8]{inputenc}
\usepackage{amssymb,bbm}
\usepackage[reqno]{amsmath}

\usepackage[colorlinks,bookmarks=false,citecolor=blue,linkcolor=red,urlcolor=blue,hyperfootnotes=true]{hyperref}

\usepackage{bm}

\begin{document}

\title{Dynamical properties of the sine-Gordon quantum spin magnet Cu-PM\\ at zero and finite temperature}

\author{Alexander C. Tiegel}
\affiliation{Institut f\"ur Theoretische Physik, Georg-August-Universit\"at G\"ottingen, 37077 G\"ottingen, Germany}

\author{Andreas Honecker}
\affiliation{Laboratoire de Physique Th\'eorique et
Mod\'elisation, CNRS UMR 8089, Universit\'e de Cergy-Pontoise, 95302 Cergy-Pontoise Cedex, France}

\author{Thomas Pruschke}
\thanks{Deceased.}
\affiliation{Institut f\"ur Theoretische Physik, Georg-August-Universit\"at G\"ottingen, 37077 G\"ottingen, Germany}

\author{Alexey Ponomaryov}
\affiliation{Dresden High Magnetic Field Laboratory (HLD), Helmholtz-Zentrum Dresden-Rossendorf,
01328 Dresden, Germany}

\author{Sergei A. Zvyagin}
\affiliation{Dresden High Magnetic Field Laboratory (HLD), Helmholtz-Zentrum Dresden-Rossendorf,
01328 Dresden, Germany}

\author{Ralf Feyerherm}
\affiliation{Helmholtz-Zentrum Berlin f\"ur Materialien und Energie GmbH, 14109 Berlin, Germany}

\author{Salvatore R. Manmana}
\affiliation{Institut f\"ur Theoretische Physik, Georg-August-Universit\"at G\"ottingen, 37077 G\"ottingen, Germany}

\date{15.03.2016} 

\begin{abstract}
The material copper pyrimidine dinitrate (Cu-PM) is a quasi-one-dimensional spin system described by the spin-1/2 $XXZ$ Heisenberg antiferromagnet with  Dzyaloshinskii-Moriya interactions. Based on numerical results obtained by the density-matrix renormalization group, exact diagonalization, and accompanying electron spin resonance (ESR) experiments we revisit the spin dynamics of this compound in an applied magnetic field. Our calculations for momentum and frequency-resolved dynamical quantities give direct access to the intensity of the elementary excitations at both zero and finite temperature. This allows us to study the system beyond the low-energy description by the quantum sine-Gordon model. We find a deviation from the Lorentz invariant dispersion for the single-soliton resonance. Furthermore, our calculations only confirm the presence of the strongest boundary bound state previously derived from a boundary sine-Gordon field theory, while composite boundary-bulk excitations have too low intensities to be observable. Upon increasing the temperature, we find a temperature-induced crossover of the soliton and the emergence of new features, such as interbreather transitions. The latter observation is confirmed by our ESR experiments on Cu-PM over a wide range of the applied field.
\end{abstract}

\maketitle

\section{Introduction}
\label{sec: intro}

The isotropic spin-1/2 Heisenberg chain with antiferromagnetic nearest-neighbor exchange coupling is a paradigmatic model for quantum magnetism.
Due to the strong enhancement of quantum fluctuations on account of its low dimensionality, there is no long-range order at zero temperature. 
The ground state is a spin singlet and its elementary excitations are spinons carrying a fractional $S=1/2$. They are unbound and interact only weakly.\cite{FADDEEV1981375}  
The dynamics of this model is governed by a gapless two-spinon continuum.\cite{Karbach_PhysRevB.55.12510,Caux_2006_4spinon} 
A uniform magnetic field makes the soft modes of the excitation spectrum incommensurate,\cite{Mueller_PhysRevB.24.1429,Stone_PhysRevLett.91.037205} but leaves the spinon continuum gapless. 
Since the Heisenberg antiferromagnetic chain is in a critical phase, even small perturbations can significantly change its ground-state physics. 
A typical perturbation in spin chain materials is the presence of Dzyaloshinskii-Moriya interactions caused by spin-orbit coupling and/or a staggered $g$ tensor due to alternating crystal axes. 
In such cases, an applied magnetic field $H$ induces an effective transverse staggered field $h_{\rm stag} \propto H$, which opens an energy gap $\propto H^{2/3}$.\cite{Oshikawa_1997,Affleck_1999}
There are several realizations of such quasi-one-dimensional materials, for example, copper benzoate\cite{Dender_1996,Dender_PhysRevLett.79.1750,Asano_2000,Asano_2003}, copper pyrimidine dinitrate ([PM$\cdot$Cu(NO$_3$)$_2\cdot$(H$_2$O)$_2$]$_n$,  PM=pyrimidine; or shortly Cu-PM),\cite{Ishida19971655, Yasui2001, Feyerherm2000, Wolter_PhysRevB.68.220406, Zvyagin_2004_PhysRevLett.93.027201, Wolter_PhysRevLett.94.057204, Zvyagin_PhysRevLett.95.017207, Wolter_PhysRevLett.94.057204, Doll_PhysRevB.75.184433, Zvyagin_2011_PhysRevB.83.060409} Yb$_4$As$_3$,\cite{Kohgi_PhysRevLett.86.2439} dimethylsulfoxide CuCl$_2$,\cite{Kenzelmann2004, Kenzelmann_2005_PhysRevB.71.094411} and KCuGaF$_6$.\cite{Umegaki_2009_PhysRevB.79.184401,Umegaki2012,Umegaki_2015}\\

The low-energy degrees of freedom of these materials can be effectively described in the framework of the quantum sine-Gordon field theory.\cite{Oshikawa_1997,Essler_PhysRevB.57.10592,Affleck_1999,Furuya_2012_PhysRevLett.109.247603} One central approximation of this approach is that the field-dependent parameters such as the coupling constant $\beta(H)$ in the sine-Gordon model need to be determined in the absence of the staggered field. The quantum sine-Gordon model is exactly solvable,\cite{Luther_PhysRevB.14.2153,Zamolodchikov1979253, Bergknoff_PhysRevD.19.3666, Korepin_1979} so that many experimental observables can be evaluated for these sine-Gordon quantum magnets. This includes static properties such as the specific heat \cite{Essler_PhysRevB.59.14376} and the field dependence of the excitation energies.\cite{Oshikawa_1997,Essler_PhysRevB.57.10592,Affleck_1999}  Very important progress has been made in the prediction of dynamical properties. Nevertheless, they are accessible only for a restricted set of wave vectors. For instance, the dynamical magnetic susceptibilities can be calculated at the antiferromagnetic wave vector $q =\pi$ using the form-factor method.\cite{Essler_PhysRevB.57.10592} 
This approach has also been used to obtain the dynamical spin structure factor for both the isotropic Heisenberg chain \cite{Essler_PhysRevB.68.064410} and anisotropic $XXZ$ Heisenberg antiferromagnets \cite{Essler_2009_PhysRevB.79.024402} in a uniform longitudinal and a transverse staggered field. 
Nevertheless, the sine-Gordon theory does not fully capture the physics of high-energy bound-spinon states observed in neutron scattering \cite{Kenzelmann_2005_PhysRevB.71.094411} and its predictions are limited to the range of small to moderate fields. 
Beyond this, the density-matrix renormalization group (DMRG)\cite{White_1992,White_1993,dmrgbook,schollwoeck2005} has provided new insights into the field dependence of a few low-lying excitations up to strong magnetic fields.\cite{Zhao_2003_PhysRevLett.90.207204, Lou_2005, Lou_2006}
Furthermore, it has been shown that DMRG calculations for the lowest excitation are in agreement with ESR experiments probing the field dependence of the excitation gap at very strong magnetic fields.\cite{Zvyagin_2011_PhysRevB.83.060409} 
For small systems, there exist a few numerical results for dynamical properties of sine-Gordon quantum magnets, see e.g., Refs.\ \onlinecite{Kenzelmann_2005_PhysRevB.71.094411} and~\onlinecite{Iitaka_2003_PhysRevLett.90.047203}. 

In the present work, we revisit these systems and present a detailed and systematic study based on DMRG and exact diagonalization (ED)\cite{Noack_2005, Sandvik_review,HoneckerWessel_2009} calculations for momentum and frequency-resolved response functions at zero and at finite temperature. 
This gives direct access to relevant dynamical quantities probed in electron spin resonance (ESR)\cite{Oshikawa_2002, Zvyagin_2012_Review} or neutron scattering and allows us to study the emergence of new features when increasing the temperature. 
In our approach, the DMRG results for the spectral functions are computed directly in the frequency domain via a Chebyshev expansion of the Green's function \cite{Weisse_2006, Holzner_2011, Braun2014, Wolf2014a} using matrix product states (MPS).\cite{Schollwoeck_2011} 
At finite temperatures this is done in the framework of a Liouville space formulation for the dynamics of the purified finite-temperature density operator, see Ref.\ \onlinecite{Tiegel_2014_PhysRevB.90.060406} and Sec.\ \ref{sec: method} below. 
This combination of methods allows us to elucidate various aspects of previous predictions and findings with a high resolution, and to make a prediction for the evolution of intensities of the spectral functions with temperature.

At $T=0$, we study the ESR resonance modes and their intensities for a wide range of the applied magnetic field and at higher frequency. 
Although the predictions by sine-Gordon field theory are in very good agreement with many of the resonance frequencies in ESR experiments, a systematic deviation for the single-soliton resonance probed in Cu-PM \cite{Zvyagin_2004_PhysRevLett.93.027201} has been observed. We are able to resolve this discrepancy with the help of our DMRG results for the dynamical spin structure factor.

As ESR experiments on Cu-PM \cite{Zvyagin_2004_PhysRevLett.93.027201, Zvyagin_PhysRevLett.95.017207, Zvyagin_2011_PhysRevB.83.060409} and $\text{KCuGaF}_\text{6}$ \cite{Umegaki_2009_PhysRevB.79.184401} also probed excitations which cannot be explained in the bulk sine-Gordon theory, Furuya and Oshikawa studied boundary and impurity effects in sine-Gordon quantum magnets.\cite{Furuya_2012_PhysRevLett.109.247603} 
Using a boundary sine-Gordon field theory approach, they found that there is only one type of boundary bound state (BBS) for the soliton, antisoliton, and the first breather, which is found below the bulk gap. The energy of this BBS is also known from DMRG calculations restricted to low-lying excitations.\cite{Lou_2005, Lou_2006} According to the boundary field theory, there are even more predictions for boundary resonances at $T=0$.
However, the exact intensities of these excitations have not been determined, yet. 
In this work we find that, except for the BBS, none of the boundary resonances is clearly observable in our DMRG results for the spectral functions.

At $T>0$, Ref.\,\onlinecite{Furuya_2012_PhysRevLett.109.247603} also expects additional thermally induced transitions between the sine-Gordon excitations. Since a few interbreather excitations were experimentally observed in $\text{KCuGaF}_\text{6}$,\cite{Umegaki_2009_PhysRevB.79.184401} one central goal of this paper is the observation of thermally-induced transitions between breathers in the material Cu-PM. 
To this end, we perform both numerical calculations and accompanying ESR experiments for a wide range of the applied magnetic field. We find that finite temperature can lead to excitations between the elementary breatherlike excitations of the sine-Gordon field theory at $T=0$ and are able to follow the evolution of their intensities with temperature. 

The paper is organized as follows. 
After presenting the effective model and its excitations in Sec.\ \ref{sec: model}, we review the relevant contributions probed in ESR experiments on sine-Gordon quantum magnets in Sec.\ \ref{sec: esr_mixing}. 
Next, the numerical methods used in this work are briefly presented in Sec.\ \ref{sec: method}. 
In Sec.\ \ref{sec: Tzero_results}, we study the zero-temperature ESR modes in the material Cu-PM and their intensities and compare them to previous  experiments. Section~\ref{sec: finiteT_results} is focused on thermally activated transitions between excited states and includes both numerical and experimental results. 
Finally, the main results and conclusions of the paper are summarized in Sec.\ \ref{sec: conclusion}.

\section{Effective model and its excitations}
\label{sec: model}
\begin{figure} 
\centering
 \includegraphics[width=0.95\columnwidth]{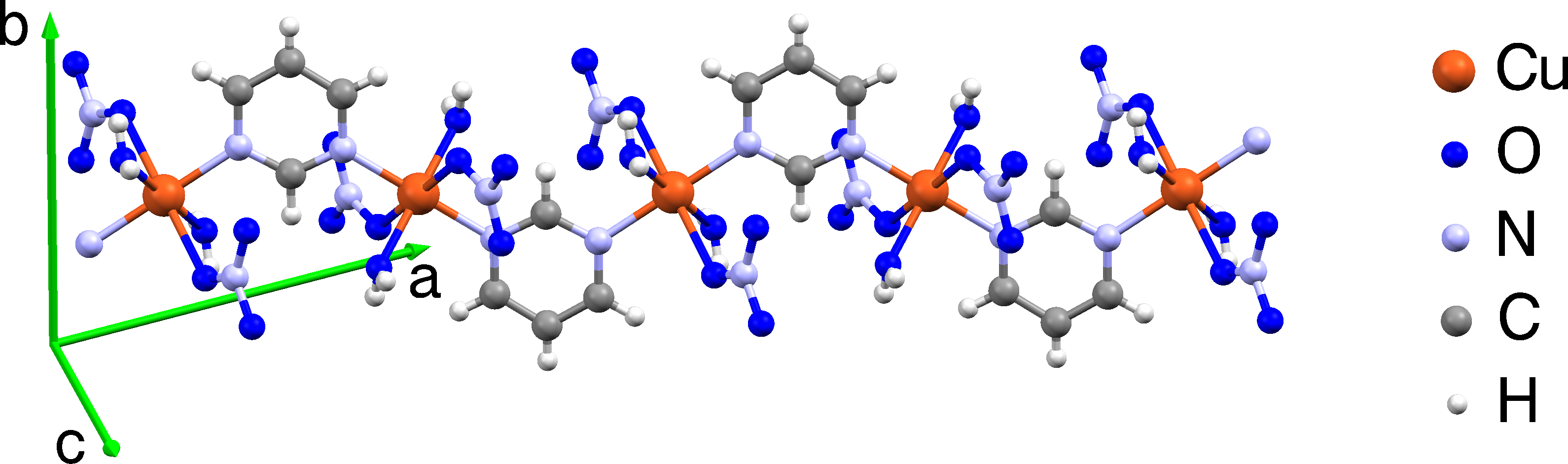} 
 \caption{ Structure of Cu-PM: chain of staggered Cu coordination octahedra linked by pyrimidine rings (running along the $a$-$c$ diagonal). The magnetic interaction $J$ is mediated by a Cu-N-C-N-Cu superexchange pathway.
} 
 \label{fig: crystal}
\end{figure}

\begin{figure}[t] 
\centering
 \includegraphics[width=0.98\columnwidth]{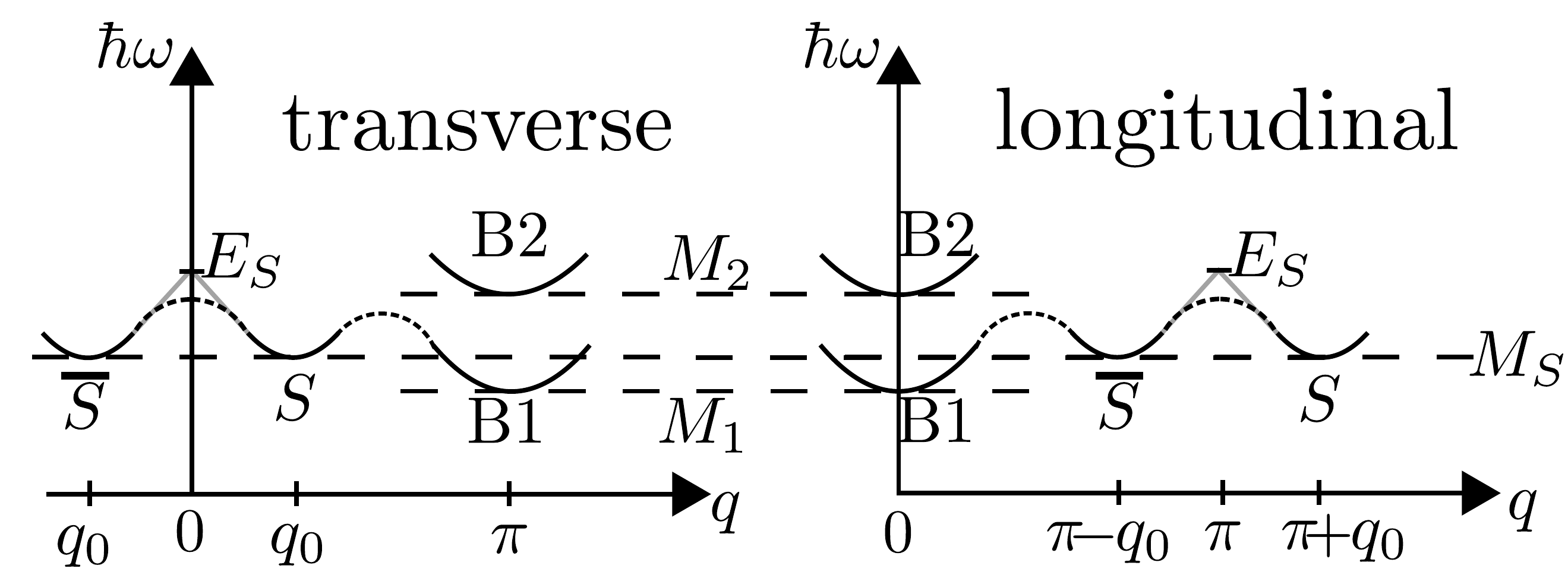} 
 \caption{Schematic sketch of the low-energy modes of the transverse (left panel) and longitudinal (right panel) dynamical susceptibility, as defined later in Eqs.\ \eqref{eq: chi_phys} and \eqref{eq: suscept}. Here B1 and B2 label the first two breathers, whereas $S$ ($\overline{S}$) denotes the (anti)soliton. As discussed in the main text, note that the modes are actually connected to each other via the dashed lines, while the field theory (thin gray lines) assumes a linear dispersion of the soliton around $q=0$, leading to a larger value of $E_S$ than observed in our DMRG results.} 
 \label{fig: low_energy_sketch}
\end{figure}

Quasi-one-dimensional spin systems such as Cu-PM\cite{Ishida19971655, Yasui2001, Feyerherm2000, Wolter_PhysRevB.68.220406, Zvyagin_2004_PhysRevLett.93.027201, Wolter_PhysRevLett.94.057204, Zvyagin_PhysRevLett.95.017207, Wolter_PhysRevLett.94.057204, Doll_PhysRevB.75.184433, Zvyagin_2011_PhysRevB.83.060409} and copper benzoate  \cite{Dender_1996,Dender_PhysRevLett.79.1750,Asano_2000,Asano_2003} possess alternating crystal axes giving rise to staggered Dzyaloshinskii-Moriya (DM) interactions and an alternating $g$ tensor. The crystal structure of Cu-PM is illustrated in Fig.\ \ref{fig: crystal}.\cite{Ishida19971655, Yasui2001, Feyerherm2000} 
For simplicity, we first neglect the effects of the anisotropic $g$ tensor and assume that a uniform magnetic field $H$ is applied along the $z$ direction, i.e., $H=h_z$. 
Furthermore, only nearest-neighbor spin exchange and DM interactions along the chain are taken into account, leading to the model Hamiltonian 
\begin{align}
  H = \sum_i \bigl[ J\, \bm S_i \cdot \bm S_{i+1} + h_z  S^z_i  + (-1)^i  \bm D \cdot  (\bm S_i \times  \bm S_{i+1})\bigr] \, .  \label{eq: microscopic_ham}
\end{align}  
The DM interactions can be eliminated\cite{Perk1976,Matsuura1977} by a staggered rotation of the spin operators about the direction of the DM vectors $\bm D_i = (-1)^i \bm D $ as long as $|\bm D | \ll J$.\cite{Oshikawa_1997,Affleck_1999}  For Cu-PM treated here, we have $J/k_B=36$ K and $D = 0.139 J$,\cite{Feyerherm2000} so that this condition is fulfilled, and we expect the effective model in Eq.\ \eqref{eq: eff_model_ham} to be valid for all strengths of the magnetic field considered. The redefinition of the spin operators generates a transverse staggered magnetic field perpendicular to both the longitudinal field and $\bm D$. Then the Hamiltonian in Eq.\ \eqref{eq: microscopic_ham} takes the simpler form 
\begin{align}
  H =  J \sum_{i} \bm S_i\cdot \bm S_{i+1}  + h_z \sum_{i} S^z_i  +  h_x \sum_{i}(-1)^i S^x_i.
    \label{eq: eff_model_ham}
\end{align}
The transverse staggered field $h_{\rm stag} = h_x$ is chosen along the $x$ direction and its strength is proportional to the uniform field, $h_x = c\, h_z$. The effect of the initially neglected anisotropy of the $g$ tensor in the Zeeman term also generates a staggered magnetic field whose transverse component we also include in the material parameter $c$, whereas the longitudinal component we assume to be very small and therefore negligible.  Note that the elimination of the DM interactions also leads to a very small exchange anisotropy which is ignored in the following. 
For $h_{\rm stag} \ll J$,  the low-energy behavior of the Hamiltonian in Eq.\ \eqref{eq: eff_model_ham} can be treated by Abelian bosonization and is given by the quantum sine-Gordon model with Lagrangian density,\cite{Oshikawa_1997,Essler_PhysRevB.57.10592,Affleck_1999,Essler_PhysRevB.59.14376}
\begin{align}
 \mathcal{L} = \frac{1}{2} (\partial_\mu \Phi)^2 + \lambda(h_{\rm stag}) \cos(\beta \Theta).     \label{eq:Lagrangian}
\end{align}
Here $\Phi$ is a boson field and $\Theta$ the corresponding dual field and the coefficient $\lambda(h_{\rm stag})$ is field dependent. Another field dependence is included in the coupling $\beta(h_z)$ which is calculated from the exact solution of the Heisenberg model in only a uniform magnetic field ($h_{\rm stag}=0$). This approximation is assumed to be justified for $h_{\rm stag}\,\ll\,h_z$. The model in Eq.\ \eqref{eq:Lagrangian} is exactly solvable\cite{Luther_PhysRevB.14.2153, Zamolodchikov1979253, Bergknoff_PhysRevD.19.3666, Korepin_1979} and the low-energy elementary excitations are known to be solitons and antisolitons which interact and propagate as robust localized quasiparticles with mass $M_S$ and charge $Q=\pm 1$. 
The soliton mass was determined for magnetic fields $h_z$ comparable to $J$ ($h_{\rm stag} \ll J$):\cite{Essler_PhysRevB.68.064410}
\begin{align}
 \frac{M_S}{J} = \frac{2v}{\sqrt \pi}\frac{\Gamma( \frac{\xi}{2})}{\Gamma( \frac{1+\xi}{2})}
  \biggl[\frac{\Gamma(\frac{1}{1+\xi})}{\Gamma(\frac{\xi}{1+\xi})} \frac{g \mu_B \pi A_x}{2 J v}\,  h_{\rm stag} \biggr]^{(1+\xi)/2}
    , \label{eq: Ms}
\end{align}
where $v$ is the dimensionless spin velocity. 
Although Equation~\eqref{eq: Ms} is exact,\cite{ZAMOLODCHIKOV_1995} the field-dependent amplitude $A_x$ for the bosonized expression of a spin operator is not known analytically. 
Therefore, both $v$ and $A_x$ were determined via DMRG for static correlation functions.\cite{Essler_PhysRevB.68.064410} 
The parameter $\xi$ is related to the field-dependent coupling $\beta$ via $\xi = \beta^2 / (8\pi - \beta^2)$.
It is important to stress that the soliton and antisoliton are found at incommensurate wave vectors $q_s= \pm q_0$ and $q_{s}= \pi \pm q_0$ as sketched in Fig.\,\ref{fig: low_energy_sketch}. 
The shift $q_0= 2 \pi m(h_z)$ is given in terms of the total magnetization per site $m(h_z)$.\cite{Affleck_1999}
As it will be reviewed in Sec.\ \ref{sec: esr_mixing}, in ESR experiments on sine-Gordon magnets only excitations at $q=0$ and $q=\pi$ can be probed.\cite{Zvyagin_2012_Review} 
Therefore, the dispersion branch linked to the soliton and antisoliton at $q_s$ is probed at these two experimentally accessible momenta. 
In order to predict the single-soliton resonance at $q=0$ resp. $q=\pi$, the field theory assumes a Lorentz invariant dispersion
\begin{align}
   E_S = \sqrt{M_S^2 + h_z^2},
\end{align}
which is sketched as a gray solid line in Fig.\,\ref{fig: low_energy_sketch}. Further elementary excitations, the breathers, consist of soliton-antisoliton bound states. 
The mass gap of the $n$th breather depends both on the soliton mass $M_S$ and on $\xi$ as 
\begin{align}
 M_n = 2 M_S \sin \left( \frac{ n \pi \xi}{2} \right). \label{eq: M_breather}
\end{align}
The number of breathers is restricted, i.e., $n = 1, 2, \ldots, \lfloor \xi^{-1}\rfloor$. 
Breather excitations do not carry any soliton charge $Q=0$.

At finite temperatures $T>0$, interbreather transitions at frequencies $M_n - M_m$ ($n>m$) are possible, and Ref.\,\onlinecite{Umegaki_2009_PhysRevB.79.184401} reports to have observed these excitations for the material KCuGaF$_6$ in pulsed-field ESR experiments, even though the measurements were performed at very low temperatures ($T/J \approx 0.005$).  Moreover, there is the field theoretical expectation for additional finite-temperature resonances at $|E_S - M_n|$.\cite{Furuya_2012_PhysRevLett.109.247603}

In the materials Cu-PM \cite{Zvyagin_2004_PhysRevLett.93.027201, Zvyagin_2011_PhysRevB.83.060409} and KCuGaF$_6$,\cite{Umegaki_2009_PhysRevB.79.184401} there are observed resonance modes which cannot be accounted for by the bulk sine-Gordon theory for an infinite system which has been discussed so far. Pursuing a boundary sine-Gordon field theory approach, Furuya and Oshikawa found that for the soliton, antisoliton, and the first breather there is only one identical boundary bound state (BBS).\cite{Furuya_2012_PhysRevLett.109.247603} This BBS is found below the bulk gap and its mass is given by
\begin{align}
  M_{\rm BBS} = M_S \sin (\pi\,\xi).
\end{align}
Furthermore, Furuya and Oshikawa argued that additional boundary resonances predicted from their theory can be assigned\cite{Furuya_2012_PhysRevLett.109.247603} to unexplained modes in the materials Cu-PM \cite{Zvyagin_2004_PhysRevLett.93.027201, Zvyagin_2011_PhysRevB.83.060409} and KCuGaF$_6$.\cite{Umegaki_2009_PhysRevB.79.184401}

\section{ESR and mixing of components}
\label{sec: esr_mixing}

In ESR\cite{Oshikawa_2002} experiments a linearly polarized electromagnetic wave is coupled to the $q=0$ component of the total spin operator $S^\alpha = \sum_i S_i^\alpha$. 
In the Faraday configuration the radiation is polarized perpendicular to the applied magnetic field in the $z$ direction, i.e., $\alpha \perp z $. 
Within linear response theory, the absorption intensity is proportional to the imaginary part $\chi''$ of the dynamical magnetic susceptibility as defined below in Eq.\ \eqref{eq: suscept}:  \cite{Pruschke_FKT2}
\begin{align}
 I(\omega) =  \frac{\pi f_0^2}{2} \, \omega \, \chi_{\text{phys}}^{''}(q=0,\omega)\, .
\end{align}
The amplitude of the electromagnetic wave is denoted by $f_0$, which for the calculations we set to $f_0 = 1$. 
Due to the staggered rotation of the spin operators performed to eliminate the DM interaction in Eq.\ \eqref{eq: microscopic_ham}, the dynamical susceptibility $\chi_{\text{phys}}$ relevant for the experiments has contributions from both uniform ($q=0$) and staggered ($q=\pi$) susceptibilities calculated for the effective model in Eq.\ \eqref{eq: eff_model_ham}:\cite{Furuya_2012_PhysRevLett.109.247603}
\begin{align}
      \chi''_{\mathrm{phys}}&(q=0, \omega) \sim \chi''_{+-}(q=0, \omega)
      \label{eq: chi_phys} \\
      &+ \biggl( \frac {D_z}J\biggr)^2 \chi''_{+-}(q=\pi, \omega)
 +\biggl( \frac{D_\perp}J\biggr)^2 \chi''_{zz}(q=\pi, \omega) \, .\nonumber
\end{align}
The mixing is determined by $D_z$ and $D_\perp$ which are the components of the DM vector $\bm D$ parallel and perpendicular to the external magnetic field pointing in the $z$ direction. 
Here $\chi''_{+-}$ ($\chi''_{zz}$) denotes the imaginary part of the transverse (resp. longitudinal) dynamical susceptibility. 
The definition in terms of the retarded Green's function corresponding to the momentum-resolved spin operators $S^\alpha_q$ and $S^\gamma_q$ is given by 
\begin{align}
\chi''_{\alpha \gamma}(q, \omega) = -\text{Im} \,\, G_{S^\alpha_q S^\gamma_q}(q,\omega). \label{eq: suscept}
\end{align}
The closely related dynamical spin structure factor then reads
\begin{align}
   S_{\alpha \gamma} (q,\omega) = \frac{1}{\pi} \, \frac{\chi''_{\alpha \gamma}(q, \omega)}{1-e^{- \omega/(k_B T)}}.
\end{align}
We set $k_B=1$ and $\hbar =1$ in our calculations. Moreover, it is known that the longitudinal susceptibility $\chi''_{zz}(q=\pi,\omega)$ contains the same resonances as the transverse contribution $\chi''_{+-}(q=0, \omega)$.\cite{Furuya_2012_PhysRevLett.109.247603} 
That is why we in particular consider $\chi''_{zz}(q=\pi, \omega)$ to study the corresponding excitations since their intensity is enhanced in this component of the dynamical susceptibility. 
Table~\ref{tab: bulk_boundary_excitations} gives an overview of the most relevant bulk and the predicted boundary excitations in the different components of the dynamical susceptibility at $T=0$.\cite{Furuya_2012_PhysRevLett.109.247603}

\begin{table}[t!]
    \begin{tabular}[t]{l | c | c  }
    \hline
    \hline
    \empty & bulk & boundary  \\
    \hline 
    $\chi''_{+-}(q=\pi, \omega)$ & \small{$\omega=M_n$, $M_n+M_m$}   & \small{$\omega=M_{\rm BBS}$, $M_n+M_{\rm BBS}$} \\[0.1cm] \hline
     $\chi''_{+-}(q=0, \omega)$ & \small{$\omega=E_S$, $E_S+M_n$}  &  \small{$\omega = E_n=\sqrt{M_n^2 + h_z^2} $,} \\
    \empty &  \empty & \small{$E_S+M_{\rm BBS}$, $E_n+M_{\rm BBS}$} \\ \hline \hline
    \end{tabular}
    \caption{Typical resonance modes from bulk resp.\ boundary sine-Gordon field
    theory at $T=0$.\cite{Furuya_2012_PhysRevLett.109.247603} Note that
    $\chi''_{zz}(q=\pi, \omega)$ contains the same resonances as $\chi''_{+-}(q=0, \omega)$.}
    \label{tab: bulk_boundary_excitations}
\end{table}

\section{Methods}
\label{sec: method}   

\subsection{DMRG at zero temperature}
\label{sec: method_dmrg_Tzero}   

The DMRG\cite{White_1992,White_1993, dmrgbook, schollwoeck2005} is arguably one of the most efficient algorithms for the computation of ground-states in one-dimensional quantum systems and can be formulated in terms of the underlying variational ansatz class of matrix product states (MPS).\cite{Schollwoeck_2011} Moreover, there exist various extensions for the calculation of Green's functions $G_{B C}(\omega+ i \eta)$ with respect to the operators $B$ and $C$, where $\eta>0$ denotes the broadening. The closely related zero-temperature spectral functions are of the form 
\begin{subequations}
\begin{align}
   A_{B C}^{T=0} (\omega) &= - \lim_{\eta \to 0} \frac{1}{\pi} \, \text{Im} \,\, G_{B C}(\omega+ i \eta) \nonumber \\
   &= - \lim_{\eta \to 0} \frac{1}{\pi} \text{Im}  \langle \Psi_0 | B \frac{1}{\omega + i \eta - (H - E_0 )}   C | \Psi_0  \rangle  \label{eq: spec_func_resolvent}  \\
   & = \langle \Psi_0 | \,B \, \delta(\omega - (H-E_0)) \, C \,| \Psi_0  \rangle. \label{eq: spec_func}
\end{align}
\end{subequations}
Here $| \Psi_0 \rangle$ is the ground state and $E_0$ its energy. Besides real-time evolution and a subsequent Fourier transform,\cite{Vidal_2004, White_2004, Daley_2004, Schmitteckert_2004} there exist DMRG approaches evaluating these expressions directly in the frequency domain. One possibility is to evaluate the resolvent for a given broadening $\eta>0$ by correction-vector methods.\cite{Ramasesha_1996,Kuehner_1999,Jeckelmann_2002, Weichselbaum_2009} However, each frequency has to be addressed individually in these approaches. An approximation of the spectral function over the entire frequency range can, for instance, be obtained by a continued fraction expansion (CFE) \cite{Haydock_1972, Gagliano_1987, Hallberg_1995, Dargel_2011, Dargel_2012} or a Chebyshev expansion.\cite{ Holzner_2011, Braun2014, Wolf2014a} We use the latter method hinged upon the kernel polynomial method\cite{Weisse_2006} and perform an MPS-based expansion of Eq.\ \eqref{eq: spec_func} in Chebyshev polynomials. Note that a Chebyshev expansion only grants convergence in the interval $\left[-1,1\right]$, since the Chebyshev polynomials $T_n(x)=\cos[n\arccos(x)]$ grow rapidly for $|x|>1$. Thus, we map the full many-body bandwidth $W$ of the Hamiltonian to $\left[-1,1\right]$ and work with a rescaled Hamiltonian. We adopt a linear rescaling scheme: 
\begin{align}
 H^\prime = \frac{H - E_0}{a} - W^\prime, \quad \omega^\prime = \frac{\omega}{a} - W^\prime,
\end{align}
where $a=W/(2 W')$ and the choice of $W'= 1- \epsilon/2$ with $\epsilon=0.025$ acts as a safeguard to strictly impose $\omega' \in \left[-1,1\right]$ . The most important part of the algorithm is the computation of the expansion coefficients
\begin{align}
\mu_n = \left\langle t_0 | t_n \right \rangle = \left\langle \Psi_0 \left| \, B \, T_n(H^\prime) \, C \, \right| \Psi_0 \right\rangle 
\end{align}
which are obtained via the recursion relation
\begin{align}
|t_0 \rangle = C | \Psi_0 \rangle, \, |t_1 \rangle = H^\prime | t_0 \rangle, \, |t_n \rangle = 2 \, H^\prime | t_{n-1} \rangle - |t_{n-2} \rangle.  \label{eq: recursion}
\end{align}
Then the spectral function is represented as 
\begin{align}
 A_{B C}^{T=0} (\omega) \approx \frac{2 W^\prime/W}{\pi  \sqrt{1 - \omega^{\prime2}}}\left[g_0  \mu_0 + 2 \sum\limits_{n=1}^{N-1} g_n \mu_n T_n(\omega^\prime)\right] .
\end{align}
The real numbers $g_n$ are damping factors which remove artificial oscillations occurring as consequence of the finite order $N$ of the expansion. Following Ref.\,\onlinecite{Weisse_2006}, we employ Jackson damping 
\begin{align}
g_n =  \frac{(N-n+1) \cos \frac{\pi  n}{N+1}   + \sin \frac{\pi n}{N+1} \cot \frac{\pi}{N+1} }{N+1}
\end{align}
introducing a nearly Gaussian broadening 
\begin{align}
 \eta(\omega, W) = \frac{\pi}{N} \, \frac{W}{2 W'} \, \sqrt{1- \omega^{\prime2}}, \label{eq: resolution}
\end{align}
which depends both on the frequency and the band width. At $T=0$, the support of the spectral function is mapped onto the lower part of the interval $\left[-W^\prime,W^\prime\right]$, where $\omega^\prime$ is only slightly larger than $-1$. As a consequence, the broadening varies significantly in this frequency range. In order to guarantee a uniform broadening at all frequencies, we adapt the expansion order as a function of the frequency, if not stated otherwise. The DMRG computations are performed in real arithmetics.  After each Chebyshev iteration, the new Chebyshev state $| \tilde t_n \rangle$ is variationally compressed\cite{Schollwoeck_2011} to an MPS $| t_n \rangle$ with smaller matrix dimension $m$. We control the accuracy of our  calculations by specifying $m$. This truncation leads to the compression error
\begin{align}
 \epsilon_{\rm compr} = | | \tilde t_n \rangle -  | t_n \rangle |^2.
\end{align}
If not stated otherwise, we use $m=250$ in order to evaluate the spectral line shape at $T=0$ corresponding to $\epsilon_{\rm compr} \sim 10^{-5}$. 

\subsection{DMRG at finite temperature}
\label{sec: method_dmrg_finiteT}  

The original zero-temperature method DMRG only allows for the treatment of pure states. There exist a few ways to address the challenge of mixed density operators at $T>0$. Finite-temperature spectral functions have originally been accessed by an application of the DMRG to transfer matrices.\cite{Naef_PhysRevB.60.359, Sirker_2005} A more recent DMRG method is the minimally entangled typically thermal states approach (METTS)\cite{White_PhysRevLett.102.190601} which samples over an ensemble of pure states constructed by imaginary-time evolution in order to approximate the finite-temperature state of the system. Since its introduction, METTS has been used for the calculation of static thermodynamic quantities\cite{Stoudenmire_2010, Alvarez_PhysRevB.87.245130} and  dynamical correlation functions.\cite{Binder_PhysRevB.92.125119,Bruognolo_PhysRevB.92.115105} Until today, most approaches are based on the purification of the mixed density operator, represented by matrix product states or operators and obtained by imaginary-time evolution. The dynamics are then studied by a subsequent real-time evolution.\cite{Verstraete_2004,Zwolak_2004,Feiguin_2005,Barthel_2009,Karrasch_2012_PRL,Barthel_2013,Karrasch_NJP_2013,Pizorn_2014} However, to calculate finite-temperature spectral functions in this paper, we use a recently developed MPS approach working directly in frequency space.\cite{Tiegel_2014_PhysRevB.90.060406} This method is also based on the purification trick. For this reason, we reconsider the underlying thermofield formalism\cite{thermofieldbook, Barnett_1987} and start by recapitulating that the $T>0$ dynamics of a mixed state with density operator $\rho$ is governed by the Liouville-von Neumann equation
\begin{align}
i \frac{d \rho}{d t} = \left[\hat H, \rho \right]. \label{eq: Liouville_vonNeumann}
\end{align}
If one then thinks of the density operator $\rho$ as a state vector $|\rho \rangle \rangle$ in the Liouville space of operators, Eq.\ \eqref{eq: Liouville_vonNeumann} becomes
\begin{align}
i \frac{d}{dt} | \rho \rangle\rangle = \mathcal L | \rho \rangle \rangle \label{eq: Liouville}
\end{align}
in Liouville space which is similar to the Schrödinger equation in Hilbert space. Here $\mathcal L$ represents the Liouville superoperator. According to Ref.\ \onlinecite{Barnett_1987}, each vector $|\rho \rangle \rangle$ can be identified with a pure-state wave function $| \Psi \rangle$ in a doubled Hilbert space $\mathcal{H}_P \otimes \mathcal{H}_Q$, which is the tensor product space of the physical state space $\mathcal{H}_P$ and an auxiliary space $\mathcal{H}_Q$ chosen to be isomorphic to $\mathcal{H}_P$. With this association, the Liouville superoperator serves as the Hamiltonian for the purification   $| \Psi \rangle$. For simplicity, we say that the dynamics of the pure-state wave function $| \Psi \rangle$ is governed by the Liouville operator $\mathcal L =  H_P \otimes I_Q - I_P \otimes  H_Q$, which we implement as a matrix-product operator, and where $I$ means the identity operator. The eigenvalues of the operator $\mathcal{L}$ are the differences of the eigenenergies of the Hamiltonian $H$. From this, it becomes evident that a Liouville space formulation is natural for the treatment of finite-temperature dynamics and the expression for a spectral function at $T>0$ is simplified to
\begin{subequations}
\begin{align}
A_{B C}^{T>0} (\omega) &= \frac{1}{Z} \, \sum_{n,m} \, \text{e}^{-E_n/T} \langle \psi_m| B |\psi_n \rangle \times \nonumber \\
  &\quad \quad \times \langle  \psi_n | C | \psi_m \rangle \,\delta\left( \omega - (E_m-E_n)\right) \label{eq: ed_finiteT} \\
  &= \langle \Psi_T | \,(B_P \otimes I_Q) \, \delta(\omega - \mathcal L ) \, (C_P \otimes I_Q) \,| \Psi_T  \rangle. \label{eq: spec_func_finiteT}
\end{align}
\end{subequations}
Here $| \Psi_T \rangle \in \mathcal{H}_P \otimes \mathcal{H}_Q$ denotes the thermal state which is obtained via an imaginary time evolution starting at infinite temperature:
\begin{align}
|\Psi_T \rangle =  e^{-(H_P \otimes I_Q)/(2T)}  |\Psi_{\infty} \rangle.
\end{align}
$|\Psi_{\infty}\rangle$ is an initial state with maximal entanglement between the real and the auxiliary system, as explained in detail in Ref.\ \onlinecite{Schollwoeck_2011}. The Liouville formalism can be used to recast finite-temperature spectral functions into a form very similar (see Eq.\ \eqref{eq: spec_func_finiteT}) to the $T=0$ expression in Eq.\ \eqref{eq: spec_func}. Therefore standard numerical methods working directly in the frequency domain are inherently applicable also at $T>0$. We thus use an MPS-based expansion in Chebyshev polynomials of Eq.\ \eqref{eq: spec_func_finiteT} which was found to have higher numerical stability and better convergence properties than a CFE in Ref.\ \onlinecite{Tiegel_2014_PhysRevB.90.060406}. Very similar to the Chebyshev expansion at $T=0$ described above, we perform the recursion in Eq.\ \eqref{eq: recursion} with respect to the linearly rescaled Liouville operator and use $|\Psi_T \rangle$ as the initial state. Concerning the resolution $\eta(\omega,W)$ in Eq.\ \eqref{eq: resolution}, it is important to note that the support of a finite-temperature spectral function in the central region of the band is mapped to frequencies $\omega^\prime \ll 1$ by the linear rescaling scheme and therefore $\eta$ is only weakly frequency dependent. However, the doubled system size needed for the purification and the fact that the spectral width of the Liouvillian assumes twice the width of the Hamiltonian make it computationally more expensive to obtain the same resolution for a given system as at $T=0$. For the evaluation of finite-temperature spectral functions in Sec.\ \ref{sec: finiteT_results}, we use a maximal matrix dimension of up to $m=300$ in order to enforce a compression error of $\epsilon_{\rm compr} \lessapprox 10^{-3}$.

\subsection{Exact diagonalization}
\label{sec: method_ed}   
We also present exact diagonalization results for the dynamical correlation functions.
Here we take a different approach from Ref.\ \onlinecite{Iitaka_2003_PhysRevLett.90.047203} and
evaluate the spectral representation Eq.\,\eqref{eq: ed_finiteT} that is then subjected
to Gaussian broadening. Since this can be done to machine precision, this
is referred to as exact diagonalization (ED).\cite{Noack_2005, Sandvik_review}
In order to exploit the translation invariance of the model in Eq.\ \eqref{eq: eff_model_ham}, periodic boundary conditions (PBCs) are adopted.
Full diagonalization of the Hamiltonian was performed for $L \le 16$ already some time ago, 
\cite{Wolter_PhysRevLett.94.057204} and has been extended up to $L = 18$ in the present context.
To go to bigger systems, we use the Lanczos method\cite{Lanczos_1950,Dagotto_1994} in order to
compute low-lying eigenvectors $|\psi_n \rangle$ and their eigenvalues $E_n$. To get a large
number of eigenvectors, we employ the procedure outlined in Ref.\ \onlinecite{HoneckerWessel_2009}
in order to eliminate the ``ghosts'' that are generated during the Lanczos iteration.
Since we need to truncate the sums in Eq.\,\eqref{eq: ed_finiteT}, the results obtained
in this manner are valid only for low temperatures and small frequencies even if about 2000
(for $L=24$) or even on the order of 10\,000 terms (for $L=20$) are retained. Among the advantages of this
approach, we have access to the individual eigenvalues $E_n$ and the individual terms of the
spectral representation in Eq.\,\eqref{eq: ed_finiteT}. Furthermore, only in the
postprocessing step do we need to specify temperature $T$ and perform broadening.

\section{Zero-temperature results}
\label{sec: Tzero_results}

For our numerical study of the ESR modes and their intensities, the value of $c$ is a crucial model parameter as it determines the magnitude of the staggered transverse field $h_x = c\,h_z$.  For instance, in KCuGaF$_6$ the parameter~$c$ assumes a value of $c = 0.178$ in the direction of the maximal  staggered magnetization.\cite{Umegaki_2009_PhysRevB.79.184401} In copper benzoate \cite{Dender_1996,Dender_PhysRevLett.79.1750,Asano_2000,Asano_2003} and dimethylsulfoxide CuCl$_2$ \cite{Kenzelmann2004, Kenzelmann_2005_PhysRevB.71.094411} the values are $c=0.065$ and $c=0.075$, respectively. 
This is a typical order of magnitude that also applies to Cu-PM. When the magnetic field is applied along the $c''$ direction, in which the maximal value of the staggered magnetization is assumed, the material parameters $c=0.08$ and $c=0.083$ were determined for Cu-PM.\cite{Zvyagin_2004_PhysRevLett.93.027201,Zvyagin_PhysRevLett.95.017207} 
Except for Sec.\ \ref{sec: T0_comparison_exp}, where $c=0.08$ is considered for a direct comparison to the experiments in Ref.\ \onlinecite{Zvyagin_2004_PhysRevLett.93.027201}, we adopt this value $c=0.083$ for our numerical calculations at $T=0$. 

\subsection{BBS and breather excitations}

\begin{figure}[t] 
\centering
 \includegraphics[width=0.95\columnwidth]{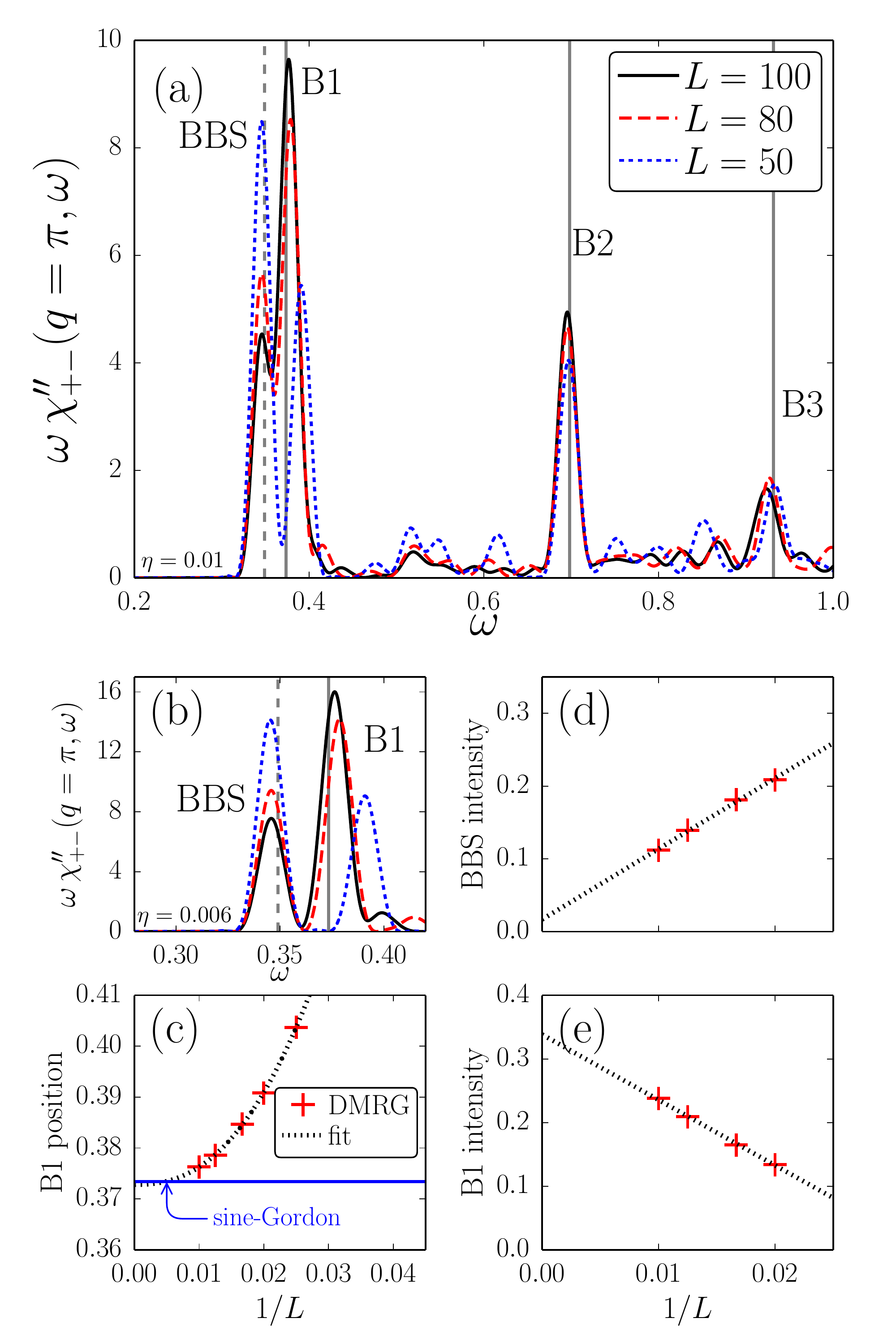}
 \caption{(Color online) Zero-temperature DMRG results for the transverse staggered contribution to the ESR intensity at $h_z=1$ ($J=1$, $c=0.083$) complemented with values from the sine-Gordon theory (vertical lines): (a) Observation of the BBS and three breathers B1--B3 in the intensity $\sim \omega \chi''_{+-}(q=\pi, \omega)$ for various system sizes $L$ and a Gaussian broadening of $\eta=0.01$. (b) Zoom-in on BBS and B1 at higher resolution ($\eta=0.006$). (c)--(e) Finite-size scaling analyses for the mass $M_1$ of B1 (c), and the integrated peak intensity of the BBS (d) as well as B1 (e).}
 \label{fig: esr_trans_spectra}
\end{figure}

We start by studying the BBS and the breather excitations which appear in the transverse staggered susceptibility $\chi''_{+-}(q=\pi, \omega)$. 
Figure~\ref{fig: esr_trans_spectra}(a) shows our DMRG results for the related ESR intensity for $h_z=1$ and $c=0.083$. 
The first three breather excitations B1--B3 are clearly observed for different system sizes $L$. 
Since the DMRG calculations are performed for open boundary conditions (OBCs), the BBS observed in earlier DMRG studies\cite{Lou_2005, Lou_2006} and finally identified as such in a boundary sine-Gordon field theory\cite{Furuya_2012_PhysRevLett.109.247603} is found slightly below B1 in the bulk gap. 
The masses of the elementary excitations from the sine-Gordon theory are included as vertical lines in Figs.\ \ref{fig: esr_trans_spectra}(a)--(b). 
The peak positions are in good agreement with the predictions. 
Moreover, the intensity decreases for the heavier quasiparticles B2 and B3. 
Figure~\ref{fig: esr_trans_spectra}(b) shows the finite-size dependence of the BBS and B1 at an enhanced resolution ($\eta=0.006$). 
Two interesting observations are that the position of the B1 peak seems to converge towards the field theoretical value, and that the BBS intensity decreases with increasing system size. 
To check this, we performed finite-size analyses using a fitting function of the form $f(x)= A + B \, x^\gamma$, where $x=1/L$. 
In Fig.\,\ref{fig: esr_trans_spectra}(c) the analysis confirms that the $L \to \infty$ extrapolation of the B1 peak position agrees with the field theoretical breather mass $M_1$, which is plotted as the horizontal solid line. 
Furthermore, we performed a finite-size scaling of the integrated peak intensity, which is independent of the broadening, both for the BBS in Fig.\,\ref{fig: esr_trans_spectra}(d) and for B1 in Fig.\,\ref{fig: esr_trans_spectra}(e). 
Here the best fit to the data was achieved by setting the exponent $\gamma=1$. 
The peak intensity of the BBS scales close to zero, whereas the intensity of B1 extrapolates to a finite value in the thermodynamic limit. 
Thus, the expectations for a boundary resp. bulk excitation are met.

\begin{figure}[t]
\centering
 \includegraphics[width=0.93\columnwidth]{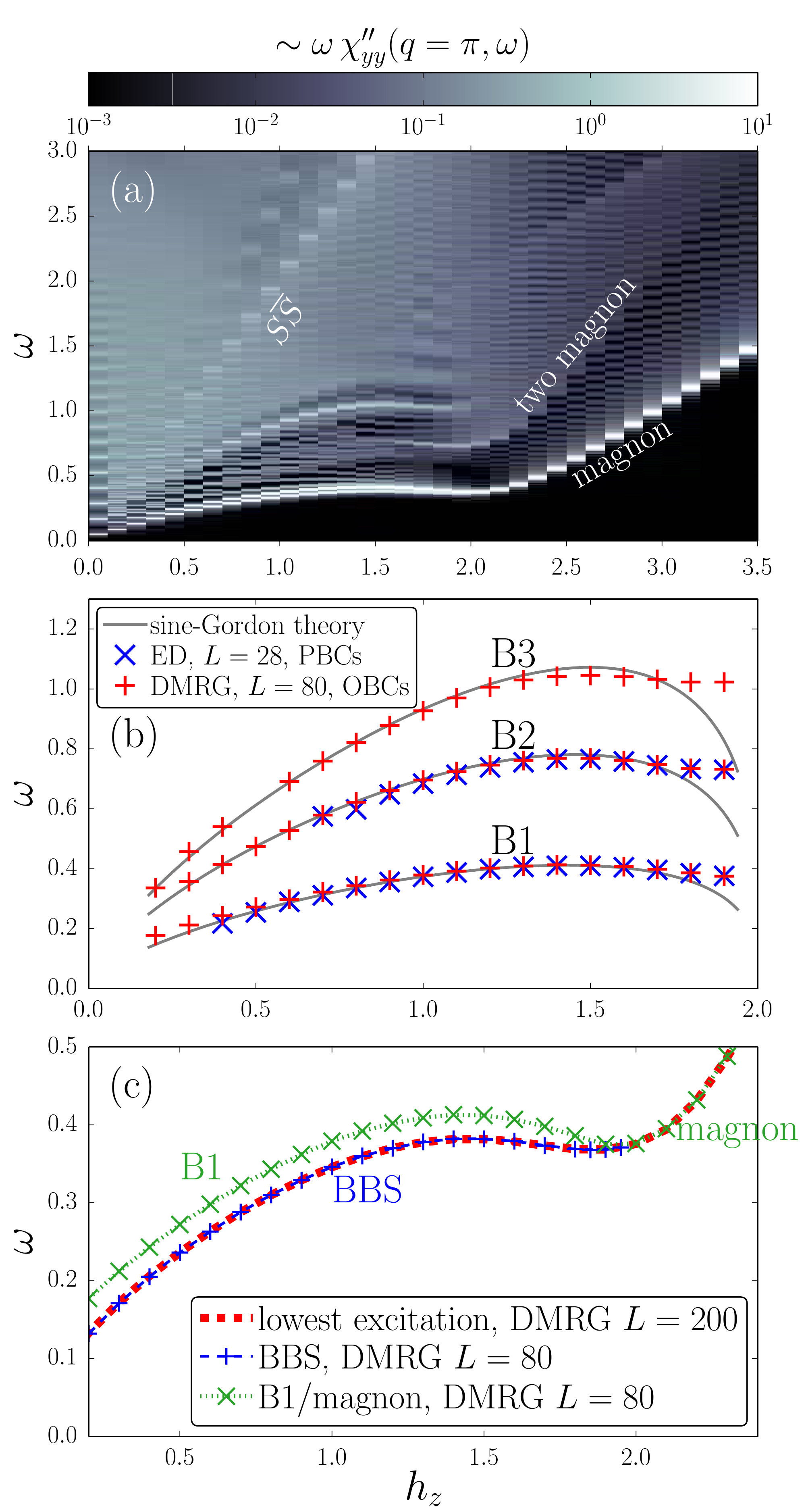} 
 \caption{(Color online) Frequency-field diagrams in the presence of an applied external field $h_z$: (a) Frequency-field plot at $T=0$ of the absorption intensity $\sim \omega \chi''_{yy}(q=\pi, \omega)$ obtained by DMRG-based Chebyshev expansions (order $N=6000$) at fixed fields $h_z \in [0,3.4]$ for a step increment of $\Delta h_z = 0.1$ and $L=80$. (b) The frequency-field dependence of the first (B1), second (B2), and third (B3) breather excitations obtained by DMRG for $L=80$ is compared to field-theoretical and ED results. (c) DMRG calculations for the frequency-field dependence of the BBS and the first breather showing the transition to magnon physics towards high fields ($J=1$ and $c=0.083$ for all panels). The DMRG results for $L=200$ are taken from Ref.\ \onlinecite{Zvyagin_2011_PhysRevB.83.060409}.}
  \label{fig: b1_b2_ffd}
\end{figure}

As a next step, the frequency-field dependence of the breather excitations B1, B2, and B3 is determined from our DMRG results for the ESR spectral function. The peak positions for B1 and B3 are obtained from the absorption intensity $\sim \omega \chi''_{yy}(q=\pi, \omega)$ for $L=80$ which is plotted as a function of the magnetic field in Fig.\,\ref{fig: b1_b2_ffd}(a).  For this calculation, $m=150$ DMRG states are kept corresponding to $\epsilon_{\rm compr}\sim 10^{-4}$ at small fields and $\epsilon_{\rm compr} < 10^{-5}$ for $h_z>1$. Since we use an MPS-based Chebyshev expansion of order $N=6000$ and the spectral width becomes larger for increasing fields, the Gaussian broadening included in these results depends both on the frequency and the field, i.e., $\eta(\omega, h_z)$. As an example, the broadening ranges from $\eta(\omega=0)=0.004$ to $\eta(\omega=3)=0.01$ at $h_z=1$ in this case. The frequency of B2 is determined from the ESR intensity $\sim \omega \chi''_{+-}(q=\pi, \omega)$, which is shown in Fig.\,\ref{fig: appendix_ffd_colorplot}(a) in  Appendix\,\ref{sec: ffd_colorplot}. In Fig.\,\ref{fig: b1_b2_ffd}(b), DMRG results for the breather resonances are compared to sine-Gordon predictions for an infinite system and exact diagonalizations (ED) of a system with $L=28$ sites. For B1, the DMRG results show deviations towards small magnetic fields, since in this field regime finite-size effects (FSEs) are gradually enhanced. However, this has been understood by the finite-size analysis in Figs.\ \ref{fig: esr_trans_spectra}(b) and~(c). 
Since the ED calculations are performed with periodic boundary conditions (PBCs), FSEs are less severe in the ED data. Interestingly, the DMRG results for B2 and B3 show a weaker finite-size dependence than for B1. In this figure one can also assess the limits of the field-theory description which is based on the limit of a small $h_x$: towards high fields, $h_z \geq 1.3$, the description by the field theory breaks down the earlier the heavier the mass of the breather excitation is. 

\begin{figure}[h]
\centering
 \includegraphics[width=0.95\columnwidth]{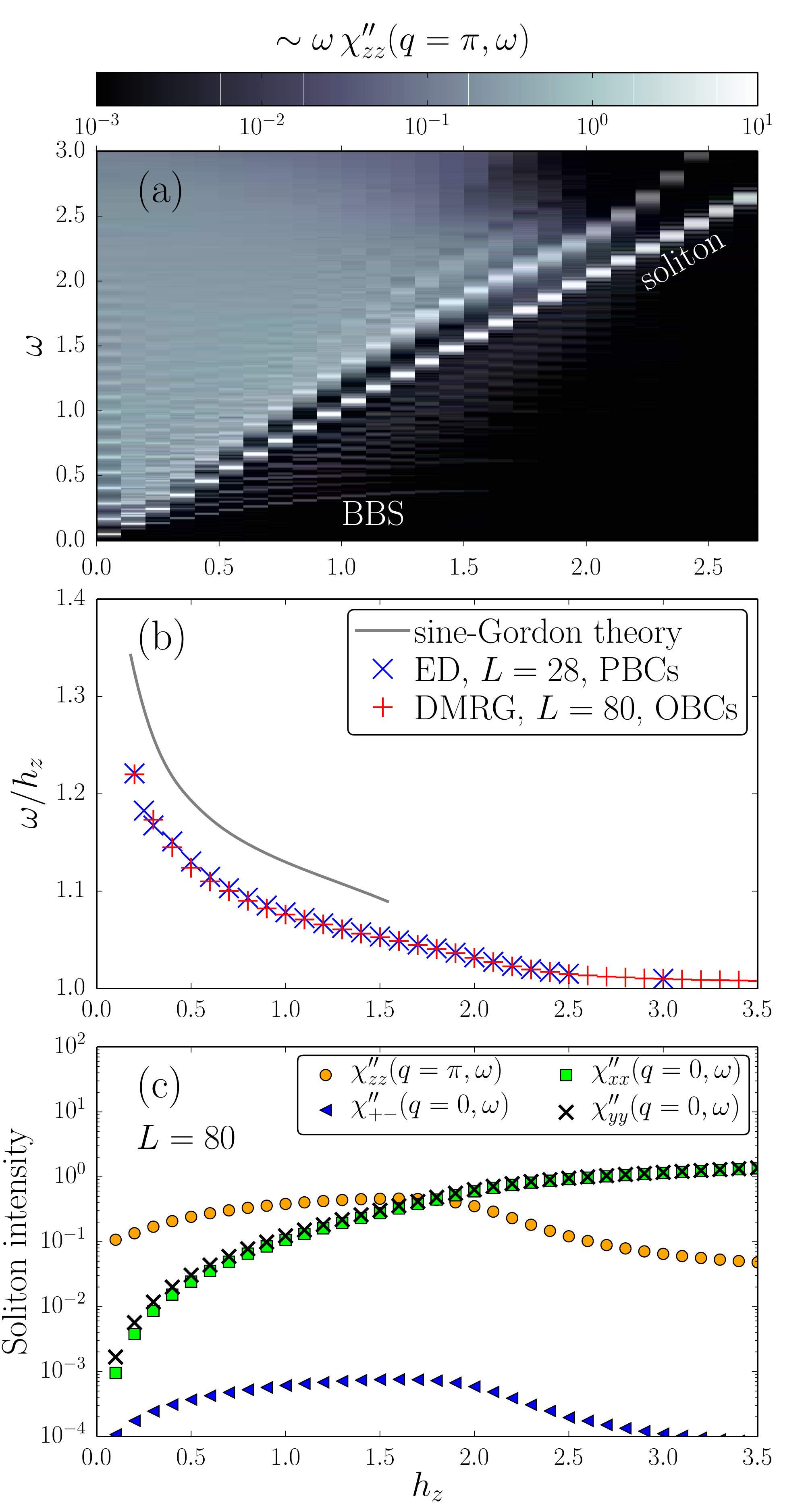}
 \caption{(Color online) Results for the soliton resonance ($J=1$, $c=0.083$): (a) Frequency-field plot of the intensity $\sim \omega \chi''_{zz}(q=\pi, \omega)$ obtained by DMRG-based Chebyshev expansions (order $N=4500$) at fixed fields $h_z \in [0,2.6]$ for a step increment of $\Delta h_z = 0.1$ and $L=80$. Here we use $m=150$ corresponding to $\epsilon_{\rm compr} \sim 10^{-4}$ at small fields. (b)~Comparison of the ratio $\omega/h_z$ for the soliton as a function of the applied field between DMRG, ED, and the sine-Gordon theory. Both numerical approaches find the soliton resonance below the field theory irrespective of the boundary conditions. (c) DMRG results for the integrated peak intensity of the soliton in different components of the dynamical susceptibility.}
 \label{fig: tzero_soliton_intensity}
\end{figure}

Moreover, it is interesting to study the evolution of the field-induced gap up to strong magnetic fields, i.e., beyond the realm of sine-Gordon physics. 
In Fig.\,\ref{fig: b1_b2_ffd}(c), we compare the frequency-field dependence of the BBS extracted from the spectral function for $L=80$ in Fig.\,\ref{fig: b1_b2_ffd}(a) to DMRG results from Ref.\ \onlinecite{Zvyagin_2011_PhysRevB.83.060409}. We note that the previously published data for the lowest excitation energy computed by a multi-target DMRG approach for $L=200$ and OBCs perfectly coincidence with our results for the BBS. Furthermore, we observe the BBS as a weak feature of the absorption intensity $\sim \omega \chi''_{zz}(q=\pi, \omega)$ in Fig.\,\ref{fig: tzero_soliton_intensity}(a) below. Besides the BBS, Fig.\,\ref{fig: b1_b2_ffd}(c) also includes results for B1. 
Interestingly, the two excitations merge into one single excitation close to the saturation field. 
In the fully spin-polarized phase at high fields, the elementary excitations are magnons and the
gap is proportional to $h_z-h_{z}^{\text{sat}}$, where $h_{z}^\text{sat}$ is the saturation field.\cite{Zvyagin_2011_PhysRevB.83.060409} Furthermore, the two-magnon continuum and in particular its lower boundary are clearly visible in Fig.\,\ref{fig: b1_b2_ffd}(a). Again, note that the resolution becomes worse towards higher fields and frequencies, since we keep the expansion order fixed at $N=6000$.

\begin{figure}[htpb]
\centering
 \includegraphics[width=0.95\columnwidth]{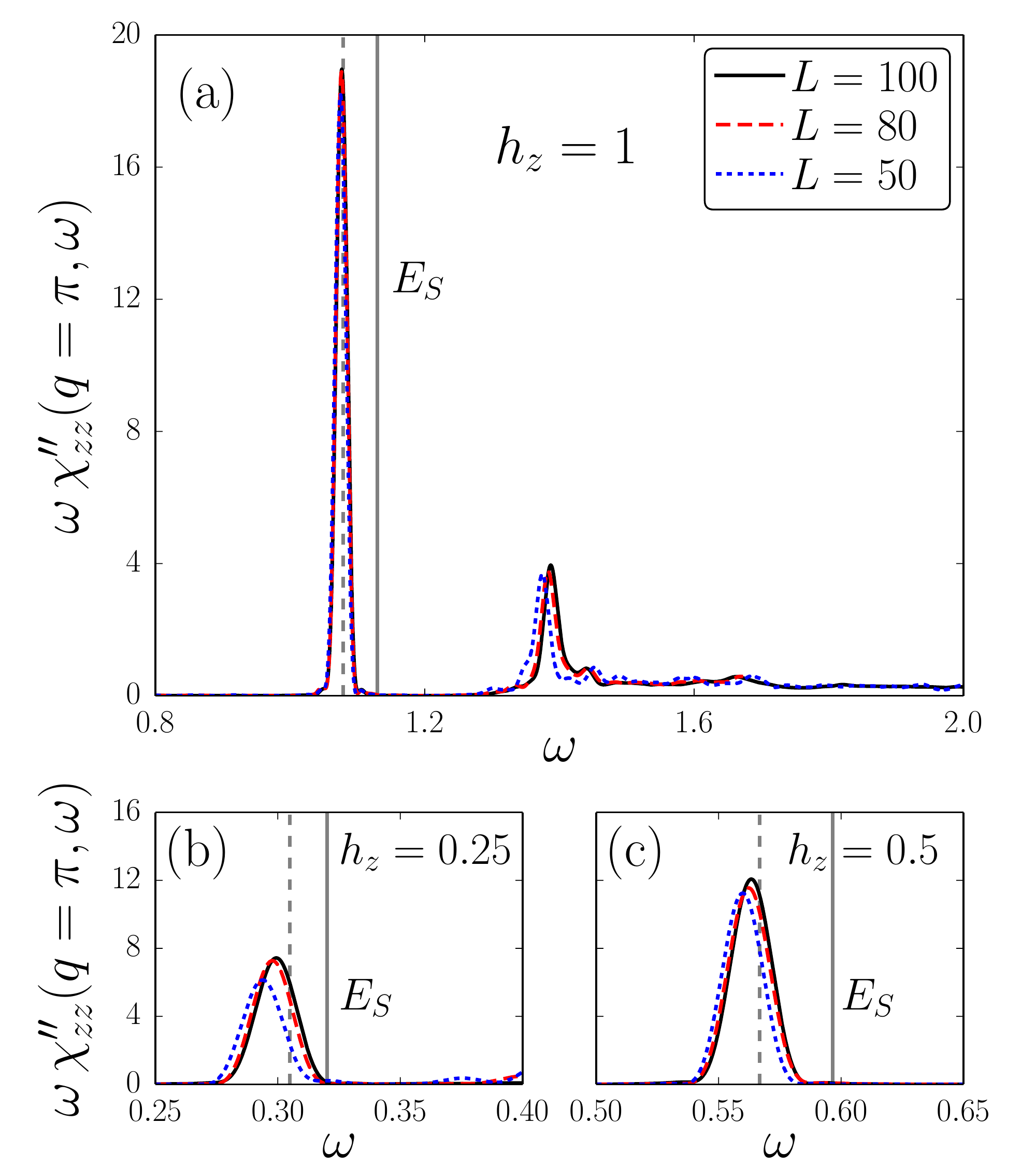}
 \caption{(Color online) Zero-temperature DMRG results for the longitudinal uniform contribution to the ESR intensity for different system sizes $L$ and three values of the applied field $h_z$: (a) $h_z=1$, (b) $h_z=0.25$, and (c)~$h_z=0.5$. We adopt $J=1$ and $c=0.083$ in all panels. The dominant peak corresponds to the single-soliton resonance and is found below the field theoretical prediction $E_S = \sqrt{M_S^2 + h_z^2}$ (solid vertical line). The dashed vertical line marks the result of the finite-size scaling.}
 \label{fig: tzero_soliton_sz_kPi}
\end{figure}

\subsection{Single-soliton resonance}

The single-soliton resonance probed in ESR experiments using the Faraday configuration mainly originates from the uniform component of the transverse susceptibility $\chi''_{+-}(q=0,\omega)$ since its presence in the longitudinal staggered susceptibility $\chi''_{zz}(q=\pi, \omega)$ is suppressed in Eq.\ \eqref{eq: chi_phys}. 
However, since the intensity of the excitation is higher in $\chi''_{zz}(q=\pi, \omega)$ by about two orders of magnitude (cf.\ Fig.\,\ref{fig: tzero_soliton_intensity}(c) below), we focus on this component for our DMRG calculations. 
Figure~\ref{fig: tzero_soliton_sz_kPi} shows our results for different system sizes $L$ and various values of the magnetic field $h_z$. 
In Fig.\,\ref{fig: tzero_soliton_sz_kPi}(a) the peak corresponding to the single-soliton resonance is the dominating feature at $h_z=1$. In addition, one observes the lower edge of a two-particle continuum at higher frequency. The extension of this continuum will be discussed in more detail in Sec.\ \ref{sec: dsf} where results for the momentum-resolved dynamical spin structure factor are presented. The $L \to \infty$ extrapolation of the single-soliton resonance is represented by the dashed vertical line in Fig.\,\ref{fig: tzero_soliton_sz_kPi}(a) and is found below the field theoretical prediction $E_S = \sqrt{M_S^2 + h_z^2}$ (solid vertical line). 
This discrepancy even persists for the smaller fields $h_z=0.25$ and~0.5 in Figs.\ \ref{fig: tzero_soliton_sz_kPi}(b) and (c), which focus on the region around the soliton resonance. 
By plotting the field dependence of the ratio $\omega / h_z$ for the soliton resonance in Fig.\,\ref{fig: tzero_soliton_intensity}(b), this discrepancy is also confirmed by ED results for $L=28$ and PBCs which are in good agreement with our DMRG calculations. 
Thus, it cannot be a boundary effect. A very similar deviation from the same theory has been observed in ESR experiments on Cu-PM\cite{Zvyagin_2004_PhysRevLett.93.027201} and will be discussed in detail in the next subsection.

Furthermore, we would like to discuss the fate of the soliton excitation after the transition into the fully spin-polarized phase. To this end, we determine the position and intensity of the single-soliton resonance from the ESR absorption $\sim \omega \chi''_{zz}(q=\pi, \omega)$ obtained by DMRG calculations for different fields in Fig.\,\ref{fig: tzero_soliton_intensity}(a) for a system of $L=80$ sites. The fact that the ratio $\omega / h_z$ approaches a value close to one for very high fields in Fig.\,\ref{fig: tzero_soliton_intensity}(b) suggests that this excitation becomes the paramagnetic line which is located at $\omega=h_z$ in standard ESR experiments and perturbed by the small staggered field here. In Fig.\,\ref{fig: tzero_soliton_intensity}(c), the integrated peak intensity from different components of the dynamical susceptibility is depicted as a function of the magnetic field. From this, we find that the highest soliton intensity appears in the longitudinal staggered susceptibility $\chi''_{zz}(q=\pi, \omega)$, whereas the intensity of the paramagnetic line at high fields is largest in the uniform transverse susceptibilities $\chi''_{xx}(q=0, \omega)$ and $\chi''_{yy}(q=0, \omega)$. An additional frequency-field diagram for the intensity  $\sim \omega \chi''_{xx}(q=0, \omega)$ is provided in Fig.\,\ref{fig: appendix_ffd_colorplot}(b) in the Appendix.

\begin{figure}[t] 
\centering
 \includegraphics[width=0.95\columnwidth]{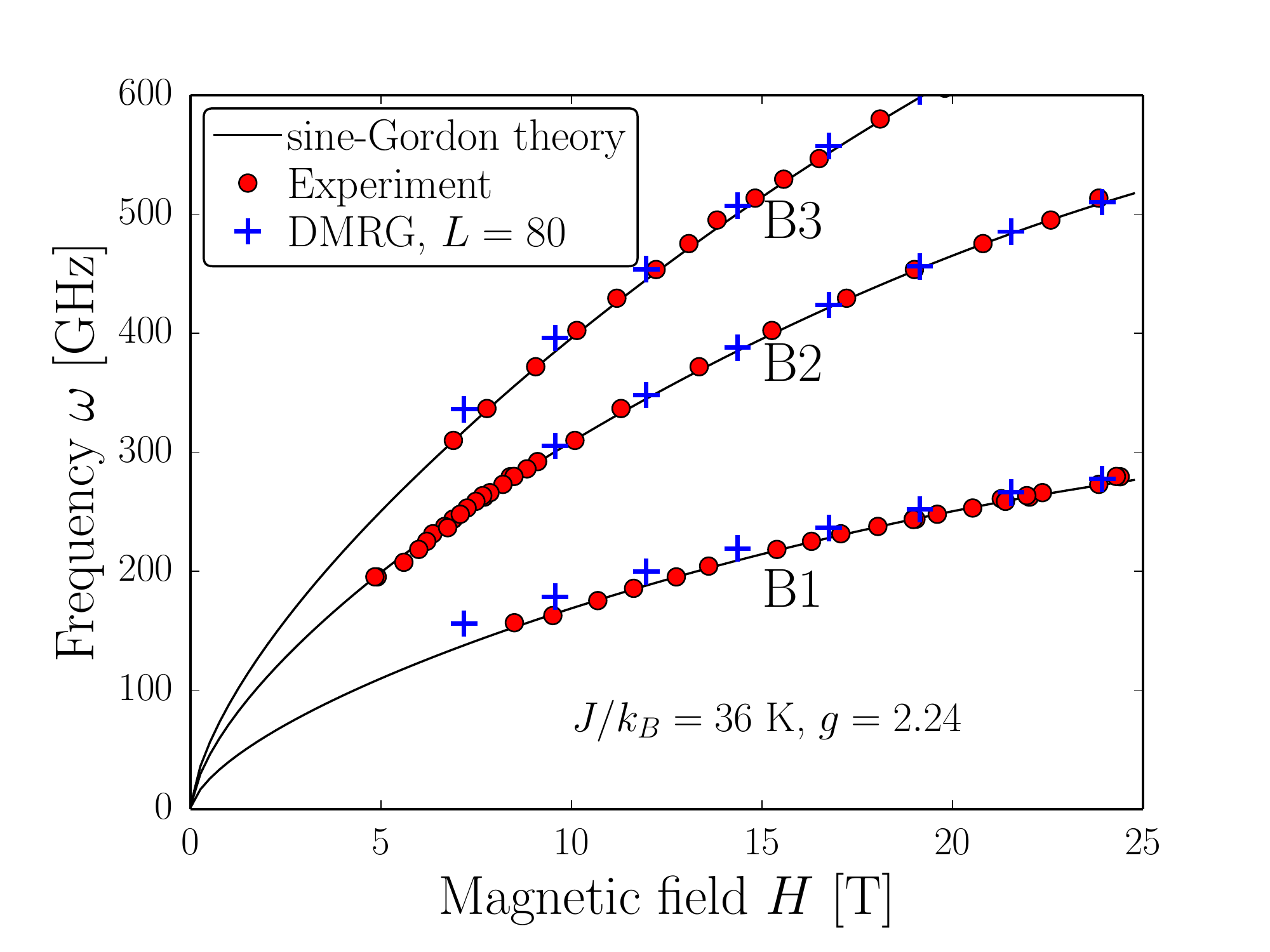} 
 \caption{(Color online) Comparison between DMRG ($L=80$), experimental, and field theoretical results for the frequency-field dependence of the breather excitations ($c=0.08$). The experimental data are taken from Ref.\ \onlinecite{Zvyagin_2004_PhysRevLett.93.027201}.} 
  \label{fig: compare_exp_br}
\end{figure}

\subsection{Comparison to the experiment}
\label{sec: T0_comparison_exp}

\begin{figure} 
\centering
 \includegraphics[width=0.95\columnwidth]{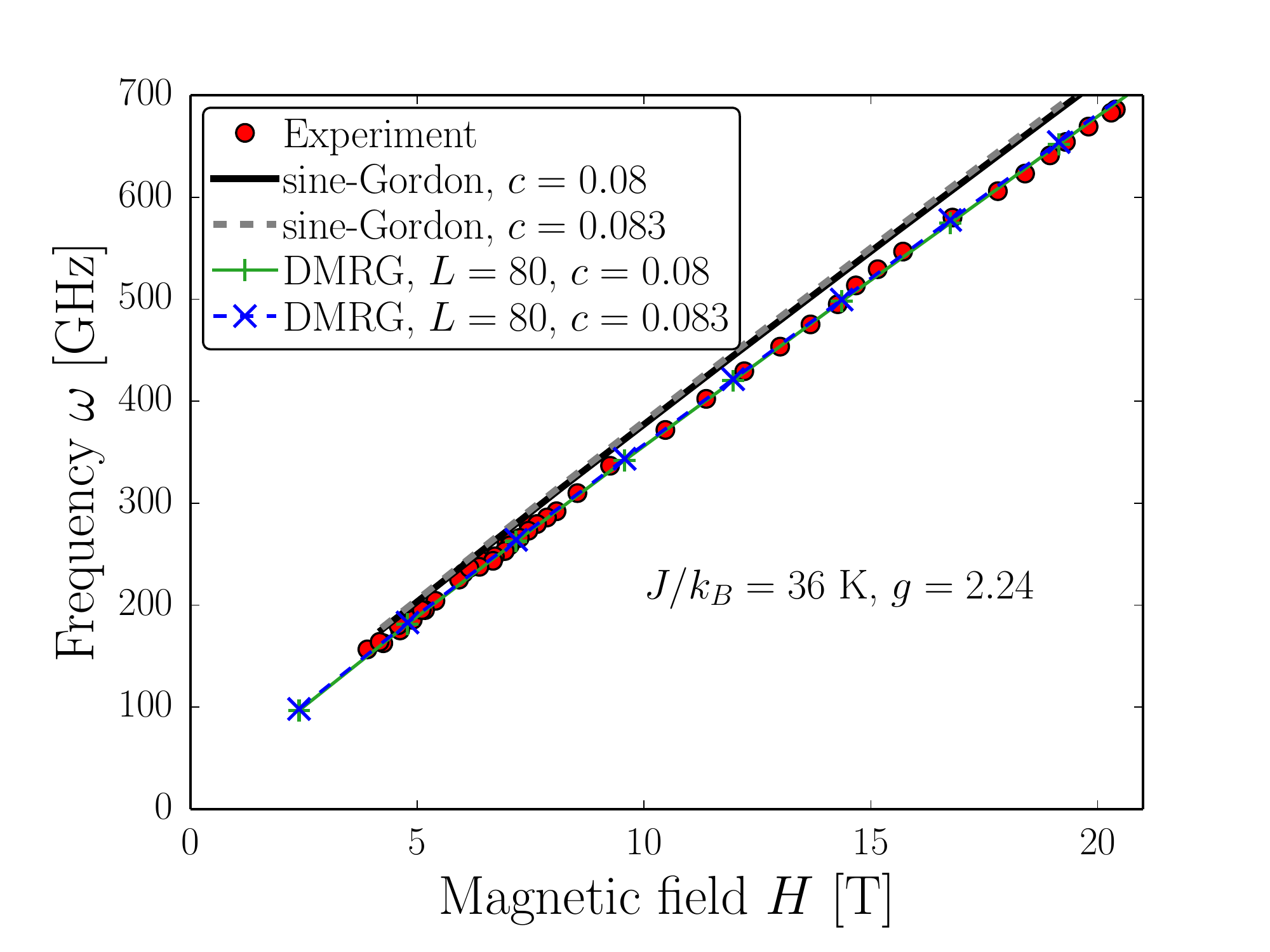} 
 \caption{(Color online) Comparison between DMRG ($L=80$), experimental, and field-theoretical results for the frequency-field dependence of the single-soliton resonance. The experimental data are taken from Ref.\ \onlinecite{Zvyagin_2004_PhysRevLett.93.027201}.} 
  \label{fig: compare_exp_sol}
\end{figure}

\begin{figure*}
\centering
 \includegraphics[width=0.9\textwidth]{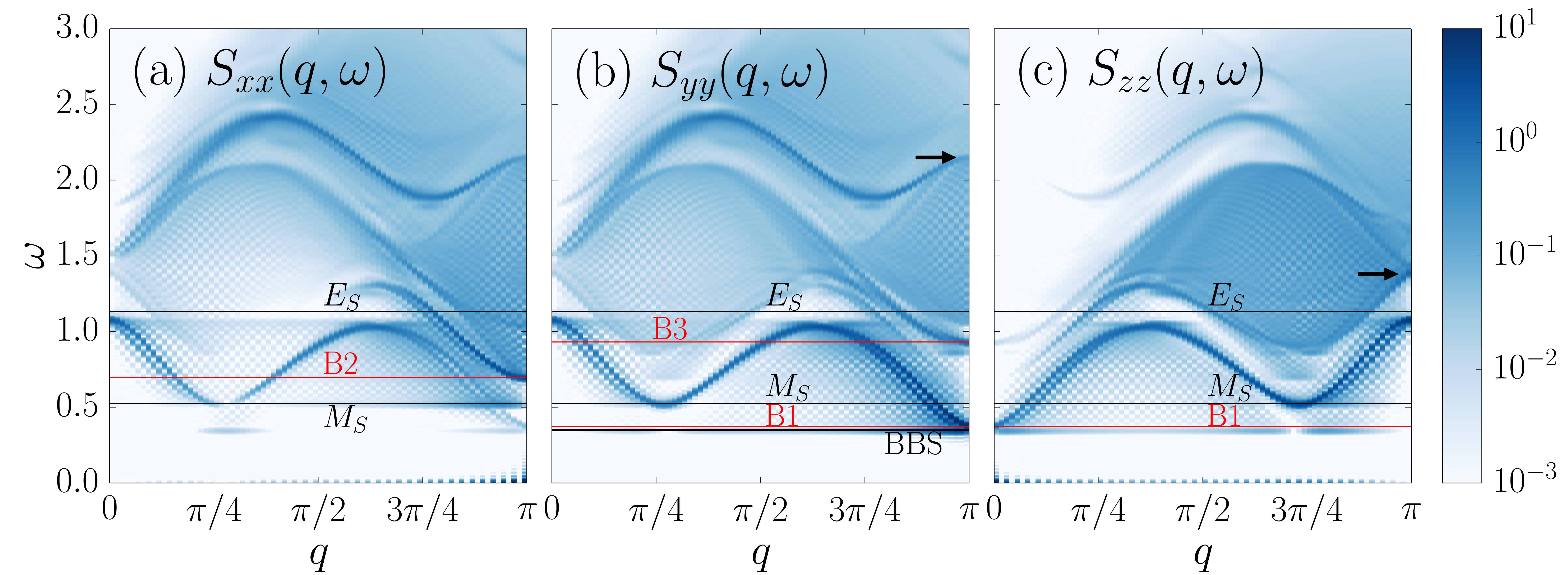}  
 \caption{(Color online) DMRG results for the $xx$, $yy$, and $zz$ components of the dynamical spin structure factor at $T=0$ for $h_z=1$ and $c=0.083$ using a uniform broadening of $\eta=0.012$ ($J=1$, $L=80$, $m=150$ and $\epsilon_{\rm compr} \sim 10^{-6}$). The horizontal solid lines represent the frequencies of the elementary excitations from the sine-Gordon theory. Most importantly, the curvature of the soliton dispersion for $q\sim 0$  in panel (a) resp. the curvature of the antisoliton for $q \sim \pi$ in panel (c) lead to a deviation from the Lorentz invariant dispersion $E_S = \sqrt{M_S^2 + h_z^2}$. The arrows at $q=\pi$ in panels (b) and (c) mark the soliton-antisoliton excitation $S\overline{S}$ and the lower edge of the continuum, respectively.}
 \label{fig: dyn_structure_factor}
\end{figure*}

In the following we want to relate our DMRG results to the experimentally determined ESR resonance modes in the material Cu-PM from Ref.\ \onlinecite{Zvyagin_2004_PhysRevLett.93.027201}. 
Since the only free parameter $c$ has been determined as  $c=0.08$ by a fit to the B1 mode in this previous experiment,\cite{Zvyagin_2004_PhysRevLett.93.027201} we adopt the value in our calculations for the comparison to the experiment. 
We start by comparing the frequency-field dependence of the breather excitations in Fig.\,\ref{fig: compare_exp_br}. 
The resonance modes B1 and B3 are extracted from the absorption $\sim \omega\,\chi''_{yy}(q=\pi, \omega)$ for $L=80$ as their intensity is higher in this component. The B2 mode is again determined from the intensity $\sim \omega\,\chi''_{+-}(q=\pi, \omega)$. 
Apart from the finite-size effects towards small magnetic fields (discussed earlier in Figs.\ \ref{fig: esr_trans_spectra}(b) and~(c)), there is good agreement between DMRG and experiment.

Next, we compare the experimental data for the frequency-field dependence of the single-soliton resonance in Cu-PM to sine-Gordon and our DMRG results for the intensity $\sim \omega\,\chi''_{zz}(q=\pi, \omega)$. 
Figure~\ref{fig: compare_exp_sol} shows this comparison for two values of the material parameter $c$ from the literature, $c=0.08$ and $c=0.083$.\cite{Zvyagin_2004_PhysRevLett.93.027201,Zvyagin_PhysRevLett.95.017207} 
One of our main results is that there is very good agreement between the DMRG results and the experiment. 
Moreover, these results are both found below the sine-Gordon field theory irrespective of the $c$-value. 
The reason for the deviation can be understood by considering the momentum dependence of the dynamical structure factor $S_{zz}(q,\omega)= \chi''_{zz}(q, \omega)/\pi$ in Fig.\,\ref{fig: dyn_structure_factor}(c).
For $q \sim \pi$, the dominating feature is the dispersion relation of the antisoliton, which clearly exhibits curvature. 
This curvature comes from irrelevant operators which break Lorentz invariance.\cite{Oshikawa_2002}
Therefore, the Lorentz invariant dispersion $E_S = \sqrt{M_S^2 + h_z^2}$ used by the field theory is not perfectly suitable for describing the single-soliton resonances appearing at $q=\pi$ in $\chi''_{zz}(q, \omega)$ and at $q=0$ in $\chi''_{+-}(q, \omega)$.

Moreover, it is important to also discuss the presence of boundary modes in Cu-PM. Due to the very low impurity concentration in the sample, the BBS just below B1 is not
experimentally observed.\cite{Zvyagin_2004_PhysRevLett.93.027201,Zvyagin_2011_PhysRevB.83.060409} Since we perform  DMRG calculations with open boundaries, we have clearly identified
this BBS in Figs.\ \ref{fig: esr_trans_spectra} and~\ref{fig: b1_b2_ffd}(c). According to the boundary field theory,\cite{Furuya_2012_PhysRevLett.109.247603} there should be even more boundary
resonances at $T=0$. However, none of them is clearly observable in our DMRG results for the spectral functions since their intensity is too low. Reference \onlinecite{Furuya_2012_PhysRevLett.109.247603} argues that some of these additional modes can be assigned to the unknown modes, which were observed in Cu-PM \cite{Zvyagin_2004_PhysRevLett.93.027201, Zvyagin_2011_PhysRevB.83.060409} and to similar modes observed in KCuGaF$_6$.\cite{Umegaki_2009_PhysRevB.79.184401} These unexplained resonances could previously not be accounted for in the bulk sine-Gordon theory. However, the fact that the BBS was not observed in Cu-PM while we did observe a significant weight for the chain lengths studied by us numerically indicates the Cu-PM samples to be very clean. Now the other boundary modes seem to have so low spectral weight that they are unobservable even in our computations. Thus, we conclude that these additional boundary modes are unlikely to explain the experimental U1 and U2 modes in Cu-PM.

\subsection{Dynamical spin structure factor}
\label{sec: dsf}

Up to this point, our results are obtained for the momenta $q=0$ and $q=\pi$, which are relevant for a comparison to ESR experiments (cf. Eq.\ \eqref{eq: chi_phys}). Now we want to go beyond this and study the full momentum dependence of the elementary excitations at $T=0$ for a magnetic field of magnitude $h_z=1$. Figure~\ref{fig: dyn_structure_factor} shows DMRG results for the $xx$, $yy$, and $zz$ components of the dynamical spin structure factor $S_{\alpha\alpha}(q,\omega)$ for an $L=80$ site chain. For momentum resolved quantities, we define the spin operators in $q$ space as \cite{Benthien_2004}
\begin{align}
S^\alpha_q = \sqrt{\frac{2}{L+1}} \, \sum_{i=1}^L \, \sin(q\,i) \, S^\alpha_i  
\end{align}
with respect to the quasi-momenta $q=\pi n / (L+1)$ and integers $n = 1,\ldots,L$. The transverse components of the dynamical structure factor in Figs.\ \ref{fig: dyn_structure_factor}(a) and (b) contain the soliton dispersion which assumes a minimum at the incommensurate wave vector $q = q_0$. The minimum of the antisoliton dispersion at $q = \pi - q_0 $ is a main feature of the $zz$ component shown in Fig.\,\ref{fig: dyn_structure_factor}(c). In the $yy$ ($zz$) component, the soliton continuously merges into the B1 dispersion, which has its minimum at the antiferromagnetic wave vector $q=\pi$ (resp. $q=0$). We can further identify the heavier breathers B2 and B3 at $q=\pi$ in $S_{xx}$ and $S_{yy}$. Interestingly, there is a manifestation of the BBS in all three components of the dynamical structure factor, while the most intense signal occurs in the $yy$ component in Fig.\,\ref{fig: dyn_structure_factor}(b). As expected for a localized mode, we also find that the BBS has a flat dispersion.\cite{Furuya_2012_PhysRevLett.109.247603}

Our results in Fig.\,\ref{fig: dyn_structure_factor} represent an improvement over a previous ED investigation of the dynamical structure factor for $L\sim20$ in Ref.\ \onlinecite{Kenzelmann_2005_PhysRevB.71.094411}. These ED calculations for $c=0.075$ are in agreement with neutron scattering results for the low-energy modes in dimethylsulfoxide CuCl$_2$ published in the same work. The DMRG calculations for $L=80$ in Fig.\,\ref{fig: dyn_structure_factor} provide a higher momentum and frequency resolution enabling us to resolve the multi-particle continua more clearly. As a main result, we are able to observe the curvature of the soliton dispersion for $q \sim 0$ in Fig.\,\ref{fig: dyn_structure_factor}(a) resp. the curvature of the antisoliton for $q \sim \pi$ in  Fig.\,\ref{fig: dyn_structure_factor}(c). The presence of this curvature is also implied by previous ESR experiments\cite{Zvyagin_2004_PhysRevLett.93.027201} since the single-soliton resonance is found below its field theoretical prediction $E_S = \sqrt{M_S^2 + h_z^2}$ (see Sec.\ \ref{sec: T0_comparison_exp}).

Moreover, multi-particle continua are observed at higher frequency. In particular, we analyzed the extended continuum in Fig.\,\ref{fig: dyn_structure_factor}(c) using the following consideration. Since there is a continuous one-particle dispersion $\epsilon_1(q)$ throughout the Brillouin zone, it is possible to construct the energies of the two-particle excitations at $q_1+q_2$ by $\epsilon_2(q_1+q_2) =\epsilon_1(q_1) + \epsilon_1(q_2)$. We indeed find that the boundaries of the continuum mostly coincidence with the extremal $\epsilon_2(q_1+q_2)$. Thus, the continuum corresponds to the continuous dispersion linking the first breather and the soliton.

An important resonance labeled as `$S\overline{S}$' in our DMRG calculations for the absorption intensity $\sim\omega \chi''_{yy}(q=\pi, \omega)$ in  Fig.\,\ref{fig: b1_b2_ffd}(a) is also found in the $yy$ component of the dynamical spin structure factor (see Fig.\,\ref{fig: dyn_structure_factor}(b)) where it is marked by an arrow. This feature is found at an energy of twice the single-soliton resonance and therefore consistent with a  soliton-antisoliton excitation. It represents the singularity at the upper edge of a continuum.

\section{Finite-temperature results}
\label{sec: finiteT_results}

Motivated by experimental hints of strong temperature dependencies of ESR line widths in Cu-PM,\cite{Zvyagin_2012_Review} we study the temperature effects on the ESR intensity of this material. We focus on the wave vectors $q=0$ and $q=\pi$, which are relevant for a comparison to ESR experiments (cf.\ Eq.\ \eqref{eq: chi_phys}). With increasing temperature, we investigate the redistribution of spectral weight and, in particular, the emergence of thermally induced transitions between zero-temperature excitations of the sine-Gordon theory. For the breather and interbreather excitations, there are both experimental and numerical results in Sec. \ref{sec: br_interbr}. In our numerical results, we furthermore observe a soliton-breather transition in Sec.\ \ref{sec: sol_br}. Moreover, temperature effects may also lead to a crossover between excitations. As an example, we discuss the temperature dependence of the soliton in Sec.\ \ref{sec: sol_Tdependence}.

\subsection{Breather and interbreather excitations}
\label{sec: br_interbr}

\begin{figure} 
\centering
 \includegraphics[width=0.95\columnwidth]{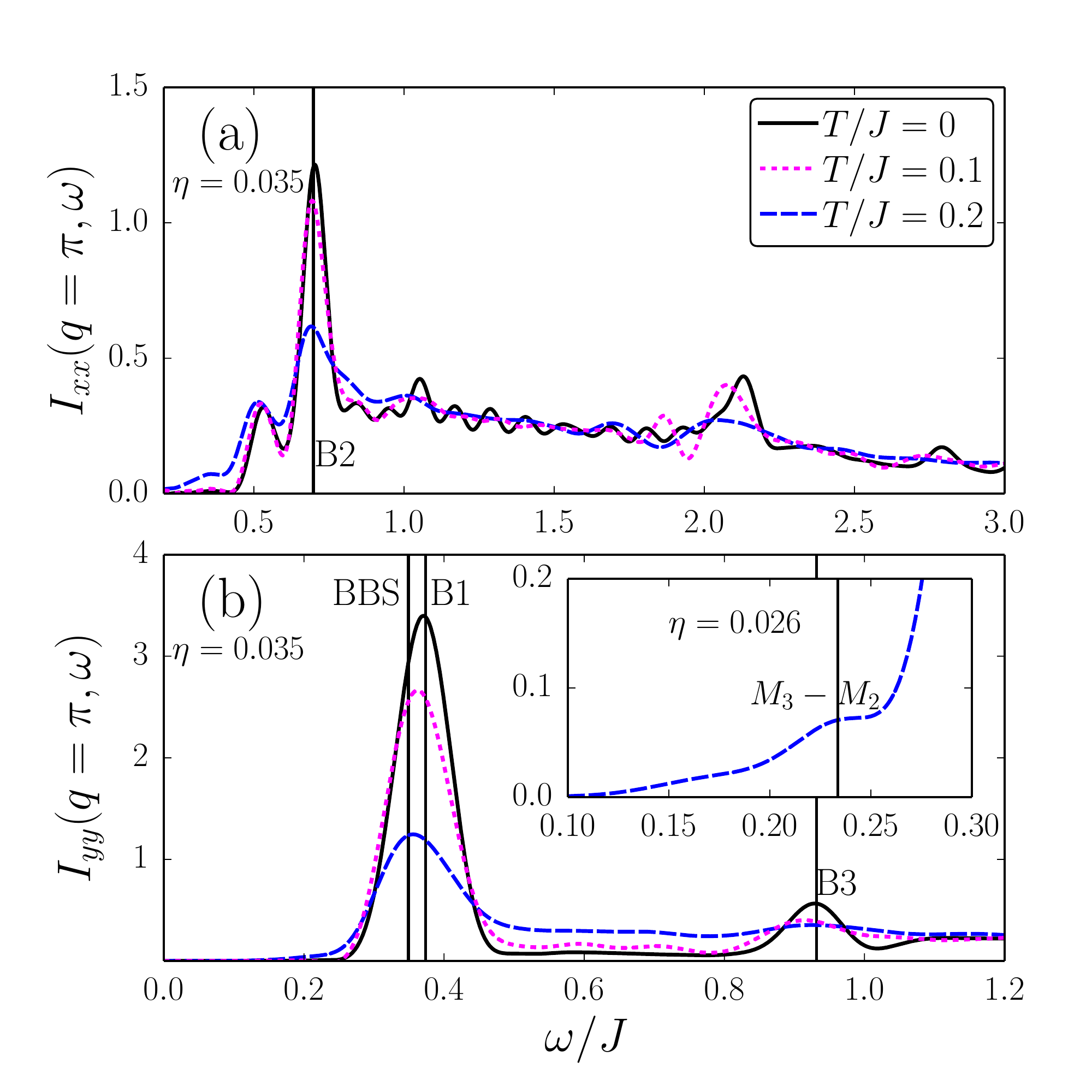}
 \caption{(Color online)  DMRG results for the temperature dependence of breather excitations in the transverse ESR intensities $I_{xx}(q=\pi,\omega)$ (a)  and $I_{yy}(q=\pi,\omega)$ (b) for a magnetic field of $h_z=1$ and $c=0.083$. The solid vertical lines mark the $T=0$ sine-Gordon predictions. Inset: DMRG calculations with enhanced resolution provide evidence for the interbreather transition at $M_3-M_2$ (solid vertical line).}
 \label{fig: mps_breather_hz1}
\end{figure}

\begin{figure} 
\centering
 \includegraphics[width=0.95\columnwidth]{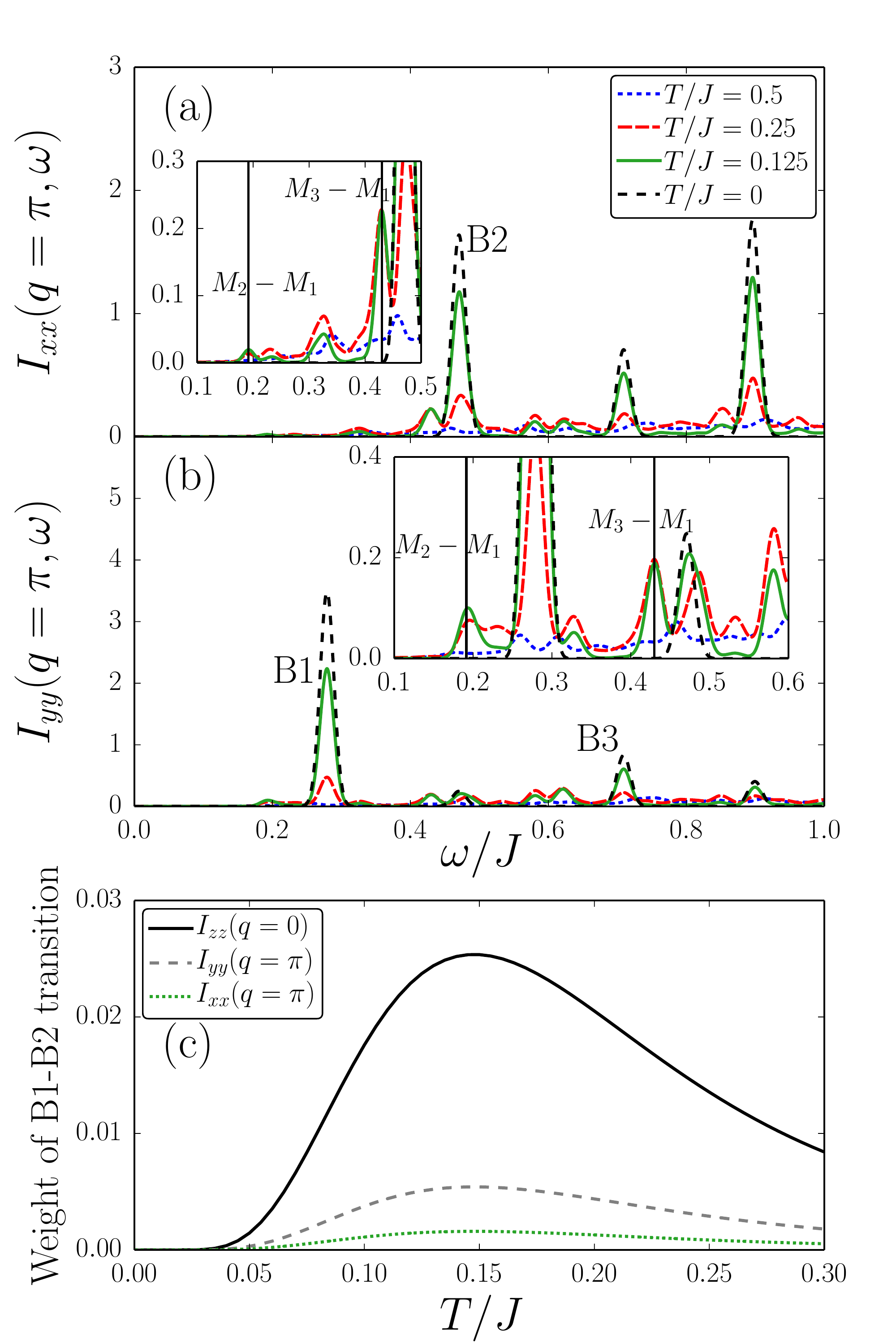}
 \caption{(Color online)  ED results for systems with PBCs addressing the temperature dependence of breather and interbreather excitations at a magnetic field of $h_z=0.6$ and $c=0.08$. The transverse ESR intensities $I_{xx}(q=\pi,\omega)$ and $I_{yy}(q=\pi,\omega)$ are shown for $\eta=0.01$ in panels (a) and (b). We show results for $L=24$ at $T \leq 0.25$ and $L=20$ at $T=0.5$. The insets focus on finite-temperature transitions at $M_n-M_m$ between breather excitations, which are marked by solid vertical lines. (c) Temperature-dependence of the spectral weight of the interbreather transition between B1 and B2 calculated from different components of the absorption intensity in dependence of temperature ($L=24$).}
 \label{fig: ed_interbreather_hz06_L24}
\end{figure}

\subsubsection{Numerical results}
First of all, we focus on the temperature dependence of the breathers at $h_z=1$. To this end, we study both contributions, $I_{xx}(q=\pi,\, \omega)$ and $I_{yy}(q=\pi,\, \omega)$, to the staggered transverse ESR intensity for $L=50$ in Fig.\,\ref{fig: mps_breather_hz1}. It is important to note that B2 is contained in the former component and B1 as well as B3 in the latter. These finite-temperature DMRG calculations for OBCs are obtained by a Chebyshev expansion with respect to the Liouville operator, as explained in Sec. \ref{sec: method_dmrg_finiteT}. In contrast to the purely $T=0$ results shown at a broadening of $\eta=0.01$ in Fig.\,\ref{fig: esr_trans_spectra}(a), the resolution in Fig.\,\ref{fig: mps_breather_hz1} assumes the value $\eta=0.035$ at $T>0$ and is therefore not as high as at $T=0$. This is due to the increased computational effort for purifying the thermal density matrix as well as applying the Liouville operator, whose spectral width is twice the width of the Hamiltonian and, most importantly, directly proportional to the broadening. As a consequence of this limitation, the BBS and B1 are not resolved as two separate peaks at $T=0$ in  Fig.\,\ref{fig: mps_breather_hz1}. However, the obtained resolution is high enough to see that at higher temperature the breather excitations are clearly subject to thermal broadening. In particular, B3 is only observable at $T=0$ in our DMRG computations due to this effect.

Since thermally activated interbreather transitions at frequencies $M_n - M_m$ ($n>m$) have been reported in the sine-Gordon magnet KCuGaF$_6$,\cite{Umegaki_2009_PhysRevB.79.184401} it is interesting to look for these excitations in our calculations for Cu-PM. We therefore enhance the resolution of the DMRG calculation at $T/J=0.5$ in the inset of Fig.\,\ref{fig: mps_breather_hz1}(b), which focuses on the region around the field-theoretical value for $M_3-M_2$ (solid vertical line). Very close to this frequency a weak maximum is found which we interpret as evidence for the $M_3-M_2$ interbreather transition. Unfortunately, it is not possible to observe further excitations of this type in our DMRG calculations. For instance, the observation of a possible feature at $\omega = M_2-M_1$ is obstructed by the choice of OBCs. More precisely, the large BBS intensity appears very close to this frequency. Therefore, we resort to ED calculations with PBCs for further detection of interbreather transitions, since the BBS is absent in this case. The ED results are shown for $L=20$ at $T=0.5$ and $L=24$ at $T \leq 0.25$ in Fig.\,\ref{fig: ed_interbreather_hz06_L24}.  Here the magnetic field is chosen as $h_z=0.6$, which corresponds to a field of about $14.36$~T in experiments on Cu-PM discussed further below. The vertical solid lines in the insets of Figs.\,\ref{fig: ed_interbreather_hz06_L24}(a) and (b) highlight the frequencies $M_n - M_m$ at which the excitations are expected. However, here these predictions are not determined via the sine-Gordon theory but from the finite-size positions of the breathers, as B2 and B3 are still subject to finite-size effects for $L=24$. In both the $xx$ and the $yy$ component of the ESR intensity, the transitions at $M_2-M_1$ and $M_3-M_1$ are clearly visible and match the expected frequencies. Figure \ref{fig: ed_interbreather_hz06_L24}(c) traces the spectral weight of the
B1-B2 transition as a function of temperature for $L=24$. Here, we have computed the coefficient
of the corresponding delta-function in the spectral representation
Eq.\ \eqref{eq: ed_finiteT}, i.e., the thermal occupation of the B1 mode multiplied with
the matrix elements for the transition to B2. The effect is that the temperature dependence
of all three quantities in Fig.\,\ref{fig: ed_interbreather_hz06_L24}(c) is identical, just the
matrix elements are different. We observe that the B1-B2 transition is thermally activated
at low $T$, goes through a maximum a little below $T/J=0.15$, and then decays again towards
high temperature. The latter is also reflected in panels (a) and (b) of Fig.\,\ref{fig: ed_interbreather_hz06_L24}
where one observes that
upon increasing the temperature to $T/J=0.5$, these weak-intensity features become hardly distinguishable from the finite-temperature background. Thus, we find the appearance of these transitions, which is expected from the sine-Gordon theory,\cite{Furuya_2012_PhysRevLett.109.247603} to be limited to low temperatures only. 

\subsubsection{Experimental results}

High-field ESR experiments on Cu-PM were performed using a 16 T superconducting magnet ESR spectrometer (similar to that described in Ref.\ \onlinecite{spectrometer}), equipped with VDI sources of millimeter-wave radiation (product of Virginia Diodes Inc.) and a transmission-type probe in the Faraday configuration. The field sweep rate was 0.5 T/min. The magnetic field was applied along the $c''$ direction, which is characterized by the maximal value of the staggered magnetization for Cu-PM.\cite{Feyerherm2000} High-quality single crystals of Cu-PM with typical size of $3\times3\times1~{\rm mm}^3$ were used that have been grown by slow evaporation of the equimolar aqueous solution of copper nitrate and pyrimidine.\cite{Ishida19971655, Yasui2001}

\begin{figure}
\centering
 \includegraphics[width=0.95\columnwidth]{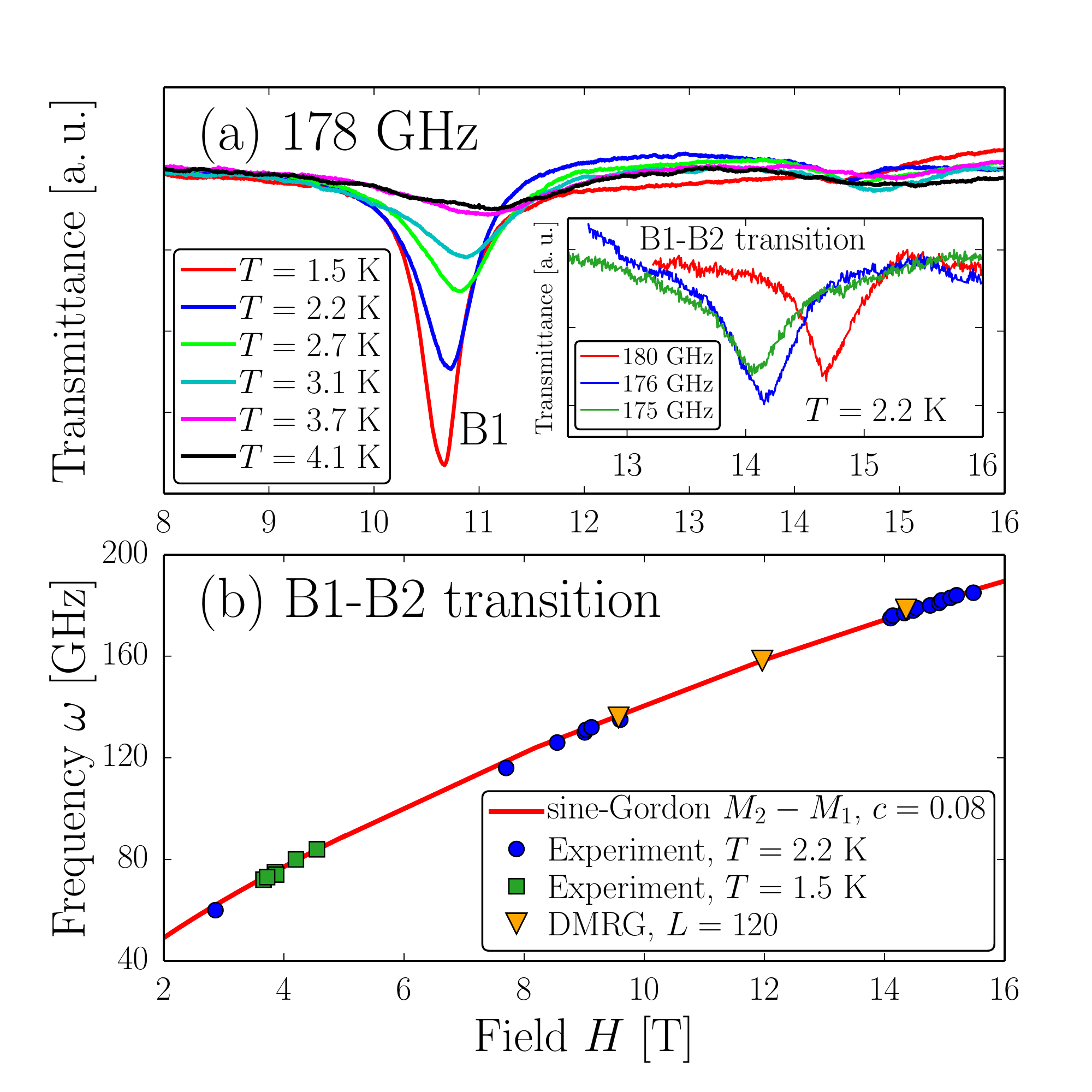}
 \caption{(Color online) (a) Temperature dependent ESR absorption intensity of the first breather B1 at 178 GHz. There is no signature of the BBS towards higher fields, but at about $H=14.5$ T the interbreather transition between B1 and B2 is observed. The inset shows this excitation for different frequencies at $T=2.2$ K. (b) Frequency-field plot of the B1-B2 interbreather transition comparing the ESR modes with $M_2-M_1$ from the sine-Gordon theory and zero-temperature DMRG calculations for $L=120$. }
 \label{fig: esr_results}
\end{figure}

Figure \ref{fig: esr_results}(a) shows the temperature dependence of ESR absorption spectra in Cu-PM measured at 178 GHz. The most prominent feature is the first breather B1 which is clearly visible up to temperatures $T \sim 3$~K. Furthermore, the measurement confirms that there is no evidence for the presence of the BBS towards higher fields. The only stable feature we observe in addition to B1 is identified as the interbreather transition at $M_2-M_1$. The inset of Fig.\ \ref{fig: esr_results}(a) shows the corresponding absorption minimum at $T=2.2$ K for a few selected frequencies. By measuring the frequency-field diagram over a broad range of the applied magnetic field, we show that there is excellent agreement with sine-Gordon and DMRG results for $M_2-M_1$ in Fig.\ \ref{fig: esr_results}(b). In the DMRG calculations, $M_1$ and $M_2$ were determined as the peak position of B1 and B2 in the absorption intensities $\sim \omega\,\chi''_{yy}(q=\pi, \omega)$ resp. $\sim \omega\,\chi''_{xx}(q=\pi, \omega)$ for large systems of $L=120$ to minimize finite-size effects at small magnetic fields.

The signals of the interbreather transition in our ESR experiments are rather weak at a temperature of $T/J \approx 0.06$. Therefore, it is surprising that the authors of Ref.\ \onlinecite{Umegaki_2009_PhysRevB.79.184401} report that the intensities of breather and interbreather excitations are of the same order in KCuGaF$_6$ at an even much lower temperature  ($T/J \approx 0.005$). However, it has to be mentioned that their ESR measurements have been performed in a pulsed magnetic field using a larger sample of size $3\times3\times3~{\rm mm}^3$. This is different from our experiments in a static magnetic field.

\subsection{Soliton-breather transition}
\label{sec: sol_br}
\begin{figure}[t] 
\centering
 \includegraphics[width=0.95\columnwidth]{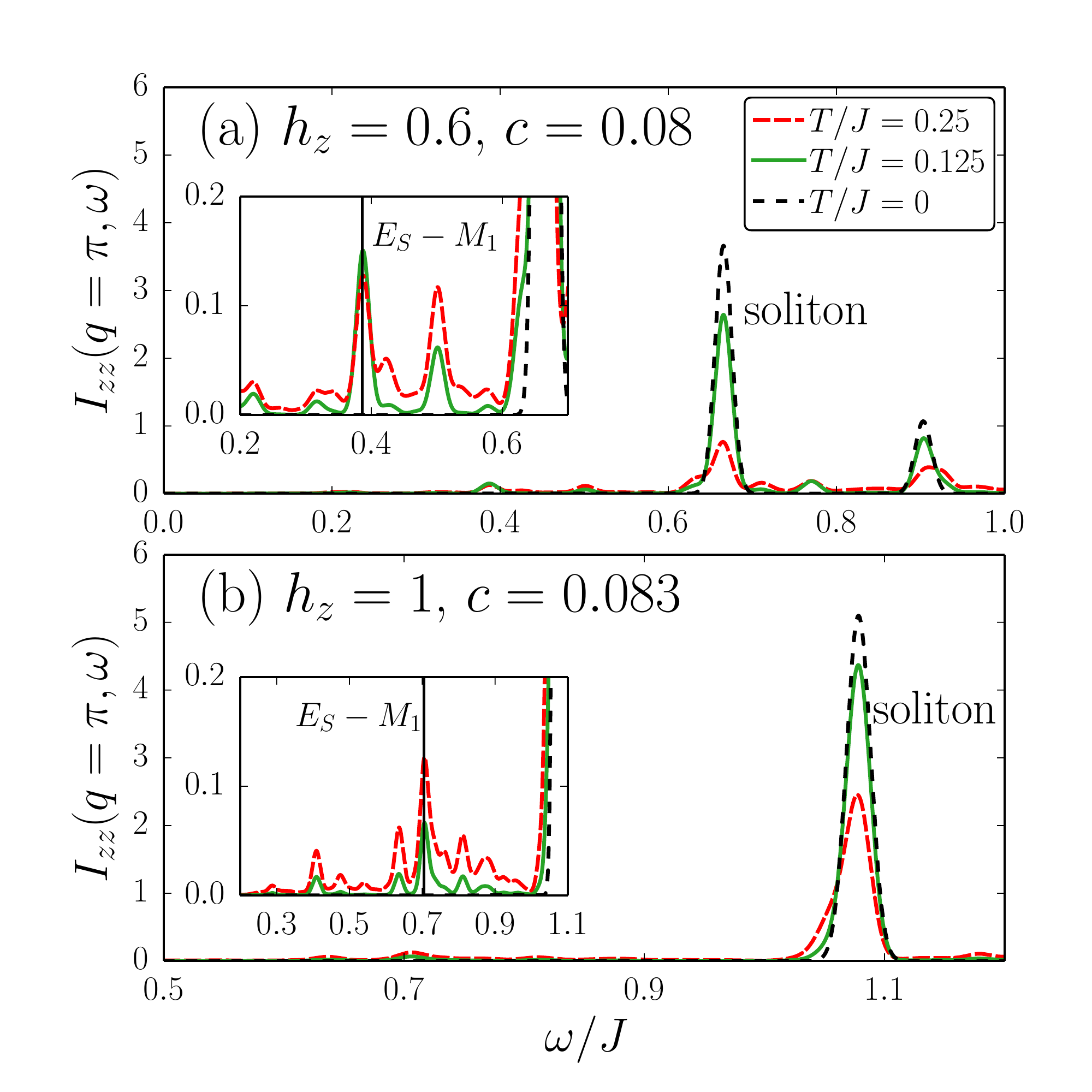}
 \caption{(Color online)  ED results for $L=24$, $\eta=0.01$, and PBCs addressing the temperature dependence of the soliton and the transition between the soliton and the first breather for different values of the magnetic field $h_z=0.6$ (a) and $h_z=1$ (b). The finite-temperature transition $E_S-M_1$ is highlighted by the solid vertical lines in the two insets.}
 \label{fig: soliton_M1_transition}
\end{figure}

Based on field-theoretical considerations, soliton-breather transitions at frequencies $|E_S - M_n|$ have been predicted to occur at finite temperature. \cite{Furuya_2012_PhysRevLett.109.247603} However, nothing has been known about the corresponding intensities so far. Our ED results for $L=24$ resolve the low-intensity transition between the soliton and the first breather at $\omega= E_S - M_1$ in the intensity $I_{zz}(q=\pi,\omega)$. The temperature dependence of this  feature is shown in the insets of Fig.\,\ref{fig: soliton_M1_transition} for magnetic fields of $h_z=0.6$ and $h_z=1$. The transition assumes its maximum intensity around a temperature of $T/J=0.25$, while it is hardly visible at $T/J=0.5$. Therefore, this feature occurs in the same temperature range as the interbreather transitions.

\begin{figure}[t]
\centering
 \includegraphics[width=0.88\columnwidth]{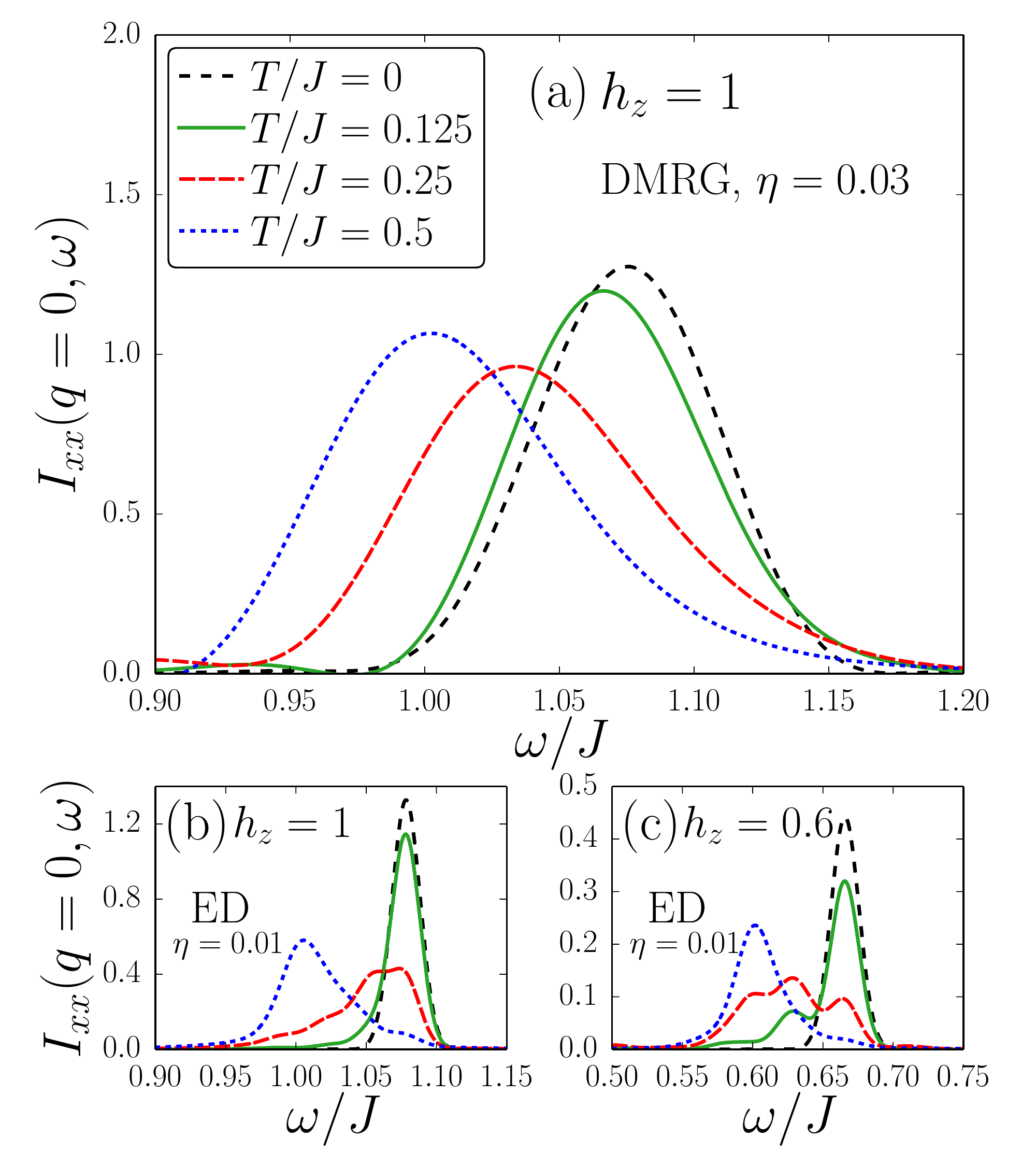}
 \caption{(Color online)  Temperature-induced crossover between the soliton at low temperature and the paramagnetic line at higher temperature in $I_{xx}(q=0,\omega)$: (a) DMRG results for $L=50$ in the presence of a uniform magnetic field $h_z=1$ and $c=0.083$ with resolution $\eta = 0.03$. (b) ED results for systems with PBCs at higher resolution $\eta = 0.01$, $h_z=1$, and $c=0.083$. (c) ED results for systems PBCs at $h_z=0.6$ and $c=0.08$ with resolution $\eta = 0.01$. The ED results are shown for $L=24$ ($T/J \leq 0.25$) and $L=20$ at $T/J=0.5$. }
 \label{fig: crossover_finiteT}
\end{figure}

\subsection{Crossover: soliton to paramagnetic line}
\label{sec: sol_Tdependence}

Next, we want to study the temperature-induced crossover between the soliton at low temperature and the paramagnetic line perturbed by the staggered field at higher temperature. For small systems, Iitaka and Ebisuzaki have presented results\cite{Iitaka_2003_PhysRevLett.90.047203} using their Boltzmann-weighted time-dependent method which is based on a random vector representation for the evaluation of the trace and a Chebyshev expansion of the Boltzmann operator. Results for lattices of $L=16$ sites have been published. Our finite-temperature DMRG approach (see Sec. \ref{sec: method_dmrg_finiteT}) allows us to revisit this feature using state-of-the-art results for larger systems with $L=50$ sites. We furthermore correct the interpretation given in Ref.\ \onlinecite{Iitaka_2003_PhysRevLett.90.047203}. Our DMRG results for different temperatures at $h_z=1$ are shown in Fig.\ \ref{fig: crossover_finiteT}(a). Figure \ref{fig: crossover_finiteT}(b) contains ED results at $h_z=1$. Here we use systems with $L=24$ ($T/J \leq 0.25$) and $L=20$ sites at $T/J = 0.5$. In Appendix\,\ref{sec: fse_ed} we show a detailed finite-size analysis for these data. ED results for $h_z=0.6$ are depicted in Fig.\ \ref{fig: crossover_finiteT}(c). For both magnitudes of the magnetic field, it is intriguing how quickly an increase in temperature redistributes spectral weight from the soliton at $T=0$ to the paramagnetic line perturbed by the staggered field at $\omega \approx h_z$ and that this crossover can be traced numerically.

\section{Conclusion}
\label{sec: conclusion}

Spectral functions for the material Cu-PM have been computed with unprecedented accuracy using DMRG and ED. At $T=0$, we studied the intensities and the frequency-field dependence of the breather excitations and the BBS found directly below the first breather, which was predicted by a boundary sine-Gordon field theory in Ref.\,\onlinecite{Furuya_2012_PhysRevLett.109.247603}. Adopting OBCs for our DMRG calculations, we could show that the BBS intensity scales to zero in the thermodynamic limit. Moreover, the first breather and the BBS merge into one single excitation close to the saturation field. Besides the BBS, Furuya and Oshikawa also predicted additional boundary modes at $T=0$ and in the case of Cu-PM assigned two of them to the previously unexplained modes U1 and U2 found for this compound.\cite{Furuya_2012_PhysRevLett.109.247603} These additional boundary resonances are not observable in our DMRG calculations for Cu-PM. Thus, we conclude that their intensities must be so low that they are unlikely to explain the U1 and U2 modes.
This conclusion is supported by the fact that in ESR experiments not even the boundary mode with the highest intensity is observed.\cite{Zvyagin_2004_PhysRevLett.93.027201}  The absence of boundary effects in our experiments can be explained by the high purity of the Cu-PM samples. A second conclusion in this context is that we have not obtained signatures for the experimental modes U1, U2, or  U3\cite{Zvyagin_2004_PhysRevLett.93.027201} in our DMRG computations, which suggests that these modes in Cu-PM may not be contained in the effective model in Eq.\ \eqref{eq: eff_model_ham}. They might occur as a consequence of further effects beyond this effective model. Possible candidates are additional anisotropies, a lattice relaxation close to the impurity, or interchain coupling.

Another main finding is that the DMRG results provide a better description for the frequency-field dependence of the single-soliton resonance in the material Cu-PM\cite{Zvyagin_2004_PhysRevLett.93.027201} than the sine-Gordon theory. The reason is that the Lorentz invariant dispersion relation $E_S = \sqrt{M_S^2 + h_z^2}$ used by the field theory does not capture the band curvature generated by the coupling of marginally irrelevant operators. This is supported by our DMRG calculations for the momentum and frequency-resolved dynamical spin structure factor in Fig.\,\ref{fig: dyn_structure_factor}.  Furthermore, recent inelastic neutron scattering experiments on KCuGaF$_6$\cite{Umegaki_2015} probed the dispersion branch along which the soliton and antisoliton are located at the incommensurate wave vectors $q_{s}= \pi \pm q_0$ as sketched in Fig.\,\ref{fig: low_energy_sketch}. We conclude that these experimental results are also compatible with the occurrence of band curvature and the single-soliton resonance at $q=\pi$.

At $T>0$, we investigated the temperature dependence of the breather and thermally activated interbreather transitions. In ED calculations with PBCs, we observed various interbreather excitations up to temperatures of about $T/J=0.5$ at both $h_z/J=0.6$ and $h_z/J=1$. In Fig.\,\ref{fig: ed_interbreather_hz06_L24}(c) the transition at $M_2-M_1$ shows maximum intensity at about $T/J=0.15$ and quickly decays upon increasing the temperature. We also observed this interbreather excitation below the first breather over a wide field range in ESR experiments on Cu-PM. The frequency-field dependence is in excellent agreement with sine-Gordon theory. Unfortunately, the open boundaries for our DMRG computations are not convenient in this case since the high BBS intensity is located very close to the frequency of the strongest interbreather excitation $M_2 -M_1$, which could therefore not be resolved. However, we found evidence for the $M_3-M_2$ transition in our DMRG data.

Finally, we revisited the single-soliton resonance which with increasing temperature crosses over to the paramagnetic line perturbed by the staggered field.  This has been studied before by Iitaka and Ebisuzaki for small systems ($L=16$) using their Boltzmann-weighted time-dependent method.\cite{Iitaka_2003_PhysRevLett.90.047203} The finite-temperature DMRG approach working directly in the frequency domain enables us to correct their interpretation by studying  larger systems with $L=50$ sites.

Concerning the effects of impurities in sine-Gordon magnets, the doping of Yb$_4$As$_3$ offers an interesting perspective for further experimental research of boundary resonances. Beyond ESR and inelastic neutron scattering on Cu-PM and other compounds, the finite-temperature MPS approach\cite{Tiegel_2014_PhysRevB.90.060406} exploiting the Liouvillian formulation of frequency-space dynamics at $T>0$ can also be applied to other systems that are, e.g., relevant in the context of ultracold gases\cite{Jin_Nature2008} and transport experiments.\\  

\textit{Note added:} We regret to announce that one of our coauthors, Prof.\ Thomas Pruschke, passed away after the submission of this article. We would like to express our gratitude for his unflagging support as a colleague and his incisive contributions as a physicist. 

\acknowledgments
We thank Shunsuke Furuya and Alexei Kolezhuk for useful discussions and acknowledge financial support of the Helmholtz Gemeinschaft via the Virtual Institute ``New states of matter and their excitations'' (Project No.\  VH-VI-521). In addition, we are grateful for the support of the HLD at HZDR, member of the European Magnetic Field Laboratory (EMFL). This work is partially supported by the Deutsche Forschungsgemeinschaft (DFG, Germany), both at HZDR (Projects No.\ ZV 6/2-1 and No.\ ZV 6/2-2) and in G\"ottingen (CRC/SFB 1073, Project B03). The crystal structure in Fig.\ \ref{fig: crystal} was visualized using the program \textit{Mercury 3.3}.\cite{Mercury}

\appendix

\section{Frequency-field plots}
\label{sec: ffd_colorplot}
\begin{figure} 
\centering
 \includegraphics[width=0.88\columnwidth]{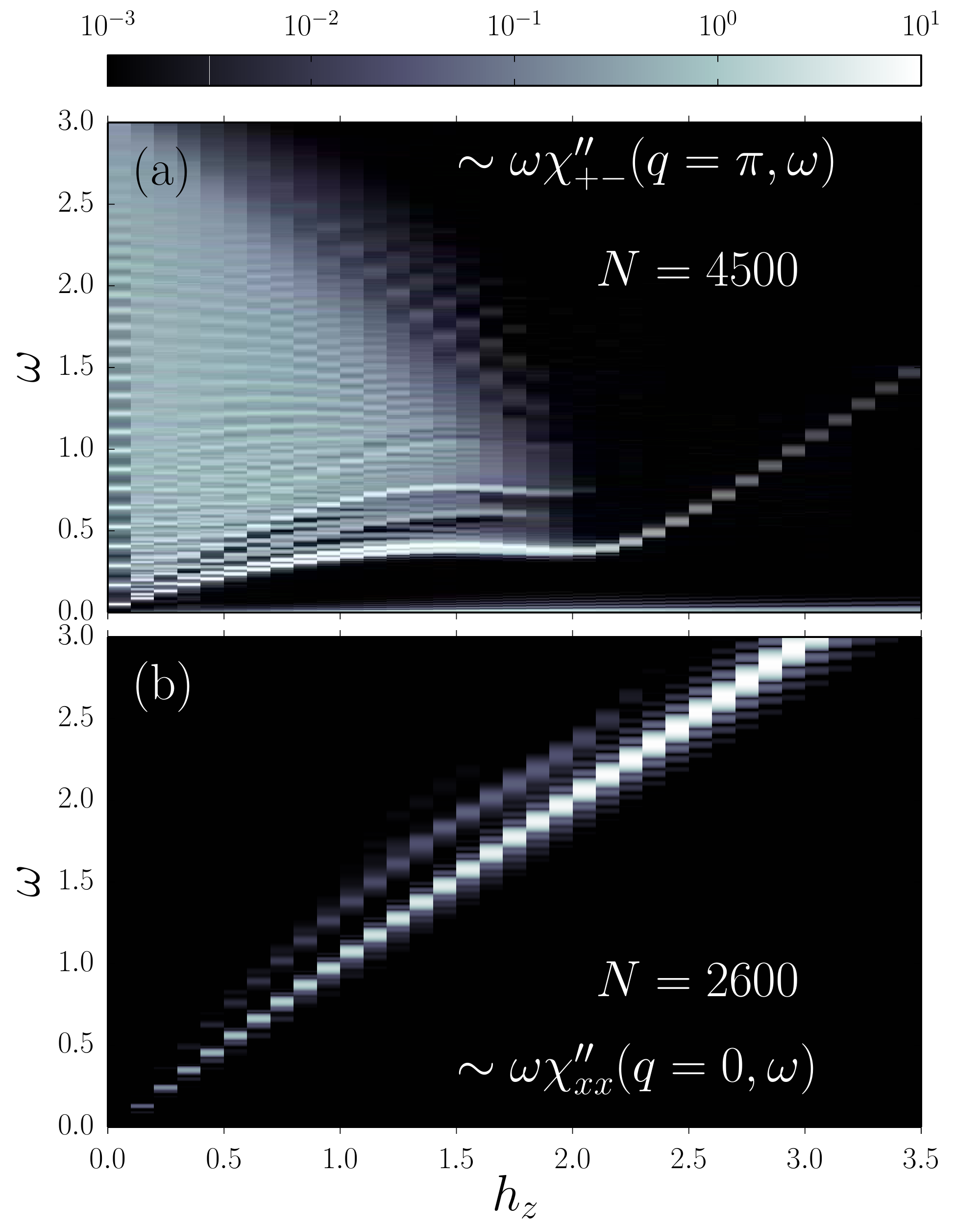}
  \caption{Frequency-field plots of the $T=0$ absorption intensities  $\sim \omega \chi''_{+-}(q=\pi, \omega)$ (a) and $\sim \omega \chi''_{xx}(q=0, \omega)$ (b) for $L=80$ and $c=0.083$. The results were obtained by DMRG-based Chebyshev expansions for fixed fields $h_z \in [0,3.4]$ for a step increment of $\Delta h_z = 0.1$.}
 \label{fig: appendix_ffd_colorplot}
\end{figure}

In this Appendix, additional frequency-field plots of the ESR absorption intensity computed by DMRG-based Chebyshev expansions at $T=0$ are provided in Fig.\,\ref{fig: appendix_ffd_colorplot}. Note the two different expansion orders $N$ and that consequently the Gaussian broadening included here depends on both $\omega$ and the spectral width $W(h_z)$.

\section{Finite-size analysis of ED results}
\begin{figure}
\centering
 \includegraphics[width=0.95\columnwidth]{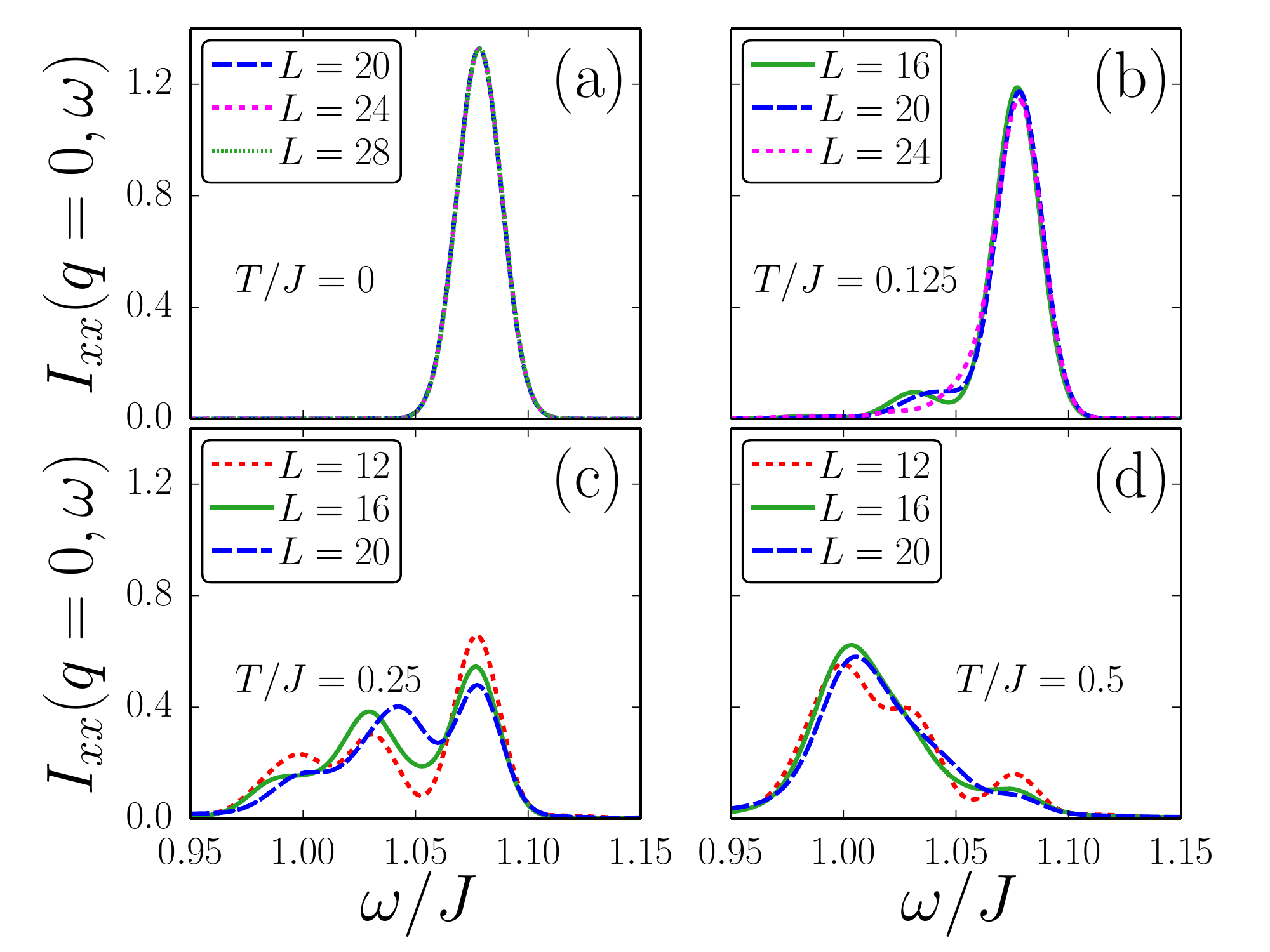}
  \caption{(Color online) Finite-size analysis of the ED results with PBCs at $h_z=1$ contained in Fig.\ \ref{fig: crossover_finiteT}(b).}
 \label{fig: fse_ed}
\end{figure}

Here we show a finite-size analysis for the ED data at $h_z=1$ shown in Fig.\,\ref{fig: crossover_finiteT}(b). To this end, we plot the ESR absorption intensity $I_{xx}(q=0,\omega)$ for different system sizes at various temperatures in Fig.\,\ref{fig: fse_ed}. We note that at $T/J=0$ the line shape resembles the thermodynamic limit for all studied system sizes, while at $T/J=0.125$ and $T/J=0.5$ the results for the largest system should be very close to those of an infinite system. Only at $T/J=0.25$, we still observe finite-size effects for $L=20$. 

\label{sec: fse_ed}

\clearpage
\newpage

\bibliography{References}

\end{document}